\documentclass[useAMS,usenatbib]{mnras}
\usepackage{newtxtext,newtxmath}
\usepackage[T1]{fontenc}
\usepackage{ae,aecompl}
\usepackage{graphicx}	
\usepackage{amsmath}	
\usepackage{amssymb}	
\usepackage{hyperref}

\newcommand\textlcsc[1]{\textsc{\MakeLowercase{#1}}}
\newcommand{\quotes}[1]{``#1''}
\defcitealias{Ucci2017}{U17}

\title[GAME]{GAME: GAlaxy Machine learning for Emission lines}
\author[G. Ucci et al.]{
G. Ucci,$^{1}$\thanks{\href{mailto:graziano.ucci@sns.it}{graziano.ucci@sns.it}},
A. Ferrara,$^{1,2}$,
A. Pallottini$^{1,3,4,5}$,
S. Gallerani$^{1}$\\
$^{1}$Scuola Normale Superiore, Piazza dei Cavalieri 7, 56126, Pisa, Italy\\
$^{2}$Kavli IPMU, The University of Tokyo, 5-1-5 Kashiwanoha, Kashiwa 277-8583, Japan\\
$^{3}$Centro Fermi, Museo Storico della Fisica e Centro Studi e Ricerche \quotes{Enrico Fermi}, Piazza del Viminale 1, Roma, 00184, Italy\\
$^{4}$Cavendish Laboratory, University of Cambridge, 19 J. J. Thomson Ave., Cambridge CB3 0HE, United Kingdom\\
$^{5}$Kavli Institute for Cosmology, University of Cambridge, Madingley Road, Cambridge CB3 0HA, UK
}

\date{Accepted XXX. Received YYY; in original form ZZZ}

\pubyear{2017}

\begin{document}
\label{firstpage}
\pagerange{\pageref{firstpage}--\pageref{lastpage}}
\maketitle

\begin{abstract}
We present an updated, optimized version of \textlcsc{GAME} (GAlaxy Machine learning for Emission lines), a code designed to infer key interstellar medium physical properties from emission line intensities of UV/optical/far infrared galaxy spectra. The improvements concern: (a) an enlarged spectral library including Pop III stars; (b) the inclusion of spectral noise in the training procedure, and (c) an accurate evaluation of uncertainties. We extensively validate the optimized code and compare its performance against empirical methods and other available emission line codes (\textlcsc{pyqz} and \textlcsc{HII-CHI-mistry}) on a sample of 62 SDSS stacked galaxy spectra and 75 observed HII regions. Very good agreement is found for metallicity. However, ionization parameters derived by \textlcsc{GAME} tend to be higher. We show that this is due to the use of too limited libraries in the other codes. The main advantages of \textlcsc{GAME} are the simultaneous use of all the measured spectral lines, and the extremely short computational times. We finally discuss the code potential and limitations.
\end{abstract}

\begin{keywords}
galaxies: ISM -- methods: data analysis -- ISM: general, lines and bands, HII regions, PDR
\end{keywords}

\section{Introduction}
Understanding the structure and physical properties of the InterStellar Medium (ISM) in galaxies, especially at high redshift, is one of the major drivers of galaxy formation studies. Measurements of key properties as gas density, column density, metallicity, ionization parameter, Habing flux, rely on galaxy spectra obtained through the most advanced telescopes (both earth-based and spaceborne) and, in particular, on their emission lines \citep{Osterbrock1989,Stasinska2007,Perez-Montero2017,method_book,Stanway2017}. However, finding diagnostics that are free of significant systematic uncertainties remains an unsolved problem \citep[and references therein]{Bresolin2016}.

Today, extensive spectroscopic studies of the redshifted UV and optical emission lines of distant galaxies ($z \gtrsim$ 2) are limited to relatively small (with respect to the local Universe) samples of star-forming galaxies \citep{Shapley2003,Erb2010,Stark2014}. The situation is even worse for higher redshifts ($z\gtrsim$ 6) where emission lines measurements are limited to few very bright galaxies \citep{Stark2017,Sobral2015} and gravitationally lensed objects \citep{Stark2015a,Stark2015b}. 
One of the main issues is that for such distant galaxies, the number of UV and optical emission lines available is almost always limited to a couple if not to a single extremely bright line. High-quality rest-frame UV and optical spectra including also fainter lines of high-$z$ galaxies, have to await for new facilities, such as the James Webb Space Telescope (JWST) and the Extremely Large Telescope (ELT). These instruments will revolutionize the current research, both in terms of quality and amount of data (estimated production rate is petabyte/year \citealt{Garofalo2017}). 

This latter aspect, in particular, will bring the astrophysical community into an era where Machine Learning (\textlcsc{ML}) algorithms and Big Data Analytics (BDA) architectures will become fundamental tools in the data mining process. This is already the case for local observations where, e.g., Integral Field Units (IFUs) are already able to provide observations of local galaxies containing tens of thousands of spaxels \citep{Cresci2017}. The development of new methods will be therefore crucial especially in the analysis of galaxy spectra which, combining the efforts from different instruments, will include faint lines arising from very extended wavelengths ranges (i.e. from UV to FIR rest-frame).

The usual emission lines diagnostics will still represent extremely useful tools, albeit they will be likely unable to reach firm conclusions as they rely on a very limited set of lines. These diagnostics are in fact based on UV/optical line ratios \citep{Pagel1979,McGaugh1991,Kewley2002,Pettini2004,Bresolin2007,Kobulnicky2004,Pilyugin2005,Bresolin2009,Marino2013,Brown2016,Pilyugin2016}. This is mostly because UV/optical Recombination Lines (RLs) emitted by the lightest elements (Hydrogen and Helium) are the strongest feature in galaxy spectra. The analogue ones emitted by heavy element can be extremely weak (up to 10$^{3-4}$ times fainter than the H$\beta$ 4861 \AA \ line), and thus very difficult to measure in very distant objects. The brightest metal lines are instead the Collisionally Excited Lines (CELs), corresponding to forbidden transitions. 

Some recent works have started to explore new diagnostics based also on FIR lines \citep{Nagao2011,Farrah2013,DeLooze2014,Vallini2015,Pallottini2015,Vallini2017,Pereira2017,Rigopoulou2018}. Facilities such as the Atacama Large Millimeter/submillimeter Array (ALMA) will allow to construct catalogues containing not only information on the UV/optical/near-IR part of the spectrum but also on the far-Infrared (FIR) lines (i.e. [CII] $\lambda$157 $\mu$m, [OI] $\lambda$63 $\mu$m, [NII] $\lambda$122 $\mu$m, [OIII] $\lambda$88 $\mu$m). Such FIR lines would perfectly complement UV/optical/near-IR lines. However, to fully exploit the information contained in the combined line set, new methods and libraries of synthetic observations are needed.

In this context, several methods in the literature have been developed that widely use ionized gas to infer galaxy physical properties from emission line intensities \citep{Stasinska2004}. A common method to measure abundances is based on the determination of the electron temperature. The electron temperature $T_e$ can be calculated from auroral to nebular emission-line ratios such as R$_{\text{O3}}$ = [OIII] $\lambda$4959+5007 / [OIII] $\lambda$4363 \citep{Perez-Montero2017} \citep{Osterbrock1989}. Alternatively, one can use the ratio of RLs of ions which show a weak dependence on $T_e$ and the electron density $n_e$ \citep{Peimbert2003,Tsamis2003,Peimbert2005,Garcia-Rojas2007,Lopez-Sanchez2007,Esteban2009,Bresolin2009b,Esteban2014,Peimbert2014,Bresolin2016b}. Given that RLs and auroral lines can be extremely weak in faint, distant or high metallicity objects, other methods which use CELs and Balmer lines have been devised to compute the ISM physical properties. These are referred to as the Strong Emission Line (SEL) methods, and they are based on the comparison of theoretical spectra from a grid of photoionization models \citep{McGaugh1991,Zaritsky1994,Kewley2002,Kobulnicky2004,Tremonti2004,Kewley2008,Dopita2016}, or on empirical calibrations obtained for samples for which a previously derived metallicity estimate via the electronic temperature method exists \citep{Pagel1979,Alloin1979,Pettini2004,Nagao2006,Maiolino2008,Nagao2011,Marino2013,Pilyugin2016,Curti2017}. In addition to these, there are codes capable to infer the abundance and ionization parameter of HII regions: \textlcsc{IZI} \citep[][based on a bayesian approach]{Blanc2015}, \textlcsc{pyqz} \citep{Dopita2013}, \textlcsc{HII-CHI-mistry} \citep{Perez2014}, \textlcsc{BOND} \citep{Asari2016}.

Within our group, we have developed \textlcsc{GAME} (GAlaxy Machine learning for Emission lines), a new fast method to reconstruct the physical properties of the ISM by using all the information represented by the emission lines intensities present in the whole available spectrum \citep[][hereafter \citetalias{Ucci2017}]{Ucci2017}. \citetalias{Ucci2017} represented a sort of \quotes{feasibility study}: our primary objective was to verify whether \textlcsc{ML} techniques can be used to predict the physical properties of galaxies. Using the AdaBoost method, we verified that the emission line intensities can effectively provide information on the state of the gas (especially for metallicity). 

As a next step, we present here the current (updated and optimized) version of the \textlcsc{GAME} code (Section \ref{sec:game}, for the workflow of the code see also Appendix \ref{sec:work}) that is based on a new library of photoionization models (50,000 synthetic spectra). In addition to the technical improvements we present also a strategy to deal with uncertainties for a \textlcsc{ML}-based code. We further implemented an approach to include noise in the library during the \textlcsc{ML} training phase, in order to apply \textlcsc{GAME} to real spectroscopic observations. Another key result concerns the treatment of the emission line degeneracy with physical properties. This is discussed in the framework of the application of \textlcsc{GAME} to study the ISM of star-forming galaxies (Section \ref{sec:application}). In Section \ref{sec:comparison} we also test the performances of \textlcsc{GAME} comparing it to other methods/codes and against a sample of HII regions with available abundances determinations. We finally discuss the potential and limitations of \textlcsc{GAME} in Section \ref{sec:discussion}.

\section{GAME}\label{sec:game}

\textlcsc{GAME} is a code that, by using as input spectral emission line intensities, infers key galaxy ISM physical properties (see \citetalias{Ucci2017} for full details). It is based on a Supervised Machine Learning algorithm called AdaBoost with Decision Trees as base learner trained with a large library of synthetic spectra. 

To generate each spectrum, we ran the photoionization code \textlcsc{CLOUDY v13.03} \citep{Cloudy}, using as input the quadruplet ($n$, $N_{H}$, $U$, $Z$) where $n$ is the total hydrogen density, $N_H$ the column density, $U$ the ionization parameter, and $Z$ the metallicity, We assume an oxygen abundance $12+\log({\rm O/H}) = 8.69$, and solar abundance ratios for all the elements\footnote{Such assumption can be relaxed, but this would require to build a dedicated library. We plan to explore the effect of peculiar abundances in future work.} \citep{Allende2001,Asplund2004,Asplund2009}. The library covers a large range of physical conditions found in the ISM, as reported in Table \ref{table:grid}. The total wavelength coverage of the synthetic spectra in the library ranges from the Ly$\alpha$ (1216 \AA) wavelength up to 1 mm. However, \textlcsc{GAME} is able to deal with any subset of emission lines within the input spectra.

A Decision Tree \citep{Breiman1984} recursively partitions the data with respect to the input feature space in branches first (i.e. the branches correspond to different regions of the input feature space), and then into an increasing number of \quotes{leaves}. Because it is possible to improve the power of many base learners into \quotes{ensemble learning methods} \citep{Dietterich2000}, we can combine many Decision Trees to make a \quotes{forest}. A common way to produce a forest of Decision Trees is the algorithm called Adaptive Boosting or Adaboost \citep{Freund1997,Drucker1997,Hastie}. AdaBoost improves the performance of a base learner, by accounting for the elements in the training set which have large prediction errors. We refer the interested reader to \citet{Schapire1990} and \citet{Drucker1997} for a detailed description of the AdaBoost algorithm\footnote{Both the implementations of AdaBoost and Decision Trees within \textlcsc{GAME} are included in the scikit-learn Python package \citep{scikit-learn}, \url{http://scikit-learn.org/.}}.

By running \textlcsc{GAME}, it is possible to infer four \quotes{default labels}: ($n$, $N_{H}$, $U$, $Z$). Besides these default labels (i.e. the physical properties used to generate the \textlcsc{CLOUDY} photoionization models, see Sec. \ref{sec:library}), in the \textlcsc{GAME} library the user can find \quotes{additional labels}, as for example the radius of the cloud $r$, the visual extinction in magnitudes $A_V$, or the FUV (6 - 13.6 eV) flux in Habing units $G/G_0$\footnote{$G_0 = 1.6\times 10^{-3} {\rm erg}\,{\rm s}^{-1}\,{\rm cm}^{-2}$ \citep{Habing1968}}. In this paper, we have chosen as additional label $G/G_0$: the GAME output is therefore a set of values for $n$, $N_{H}$, $U$, $Z$, $G/G_0$.

\subsection{Model improvements}

In this Section, we describe some new important improvements introduced here with respect to the original version of the code presented in \citetalias{Ucci2017}. These concern (a) the build-up of the spectral library, (b) the inclusion of noise in the training procedure, and (c) the evaluation of the uncertainties; they are described in the following. More technical and detailed materials can be found in the Appendixes.

\begin{table}
	\caption{Range of ISM physical properties used to construct the \textlcsc{GAME} library.}
	\centering
	\vspace{2mm}
	\begin{tabular}{c c c}
		\hline\hline
		Parameter & min value & max value\\
		\hline
		$\log(Z / Z_{\odot})$ & -3.0 & 0.5\\
		$\log(n / {\rm cm}^{-3})$ & -3.0 & 5.0\\
		$\log(U)$ & -4.0 & 3.0\\
		$\log(N_{H} / {\rm cm}^{-2}) $ & 17.0 & 23.0\\
		\hline
	\end{tabular}
	\label{table:grid}
\end{table}

\subsubsection{Library of synthetic spectra}\label{sec:library}
Concerning the library, the main improvements in the current version are:

\begin{itemize}
\item The library now contains 50,000 synthetic spectra, i.e. it is $\sim$ 65\% larger than that used in \citetalias{Ucci2017};
\item We added SEDs of Population III stars generated via the \textlcsc{YGGDRASIL} code \citep{Zackrisson2011}. We adopted the \citet{Zackrisson2011} model with a zero-metallicity population and a Kroupa IMF \citep{Kroupa2001} in the interval $0.1-100 {\rm M}_{\odot}$ based on a rescaled single stellar population from \citet{Schaerer2002}. For the star formation history we have chosen an instantaneous burst with age set to 2 Myr;
\item In addition to the graphite and silicate dust grains, we have added Polycyclic Aromatic Hydrocarbons (PAH) as these particles considerably affect EUV extinction, and are an important heating source especially in neutral regions. These contributions are modelled following \citet{Weingartner2001}. For the effect of stochastic heating we refer to the work of \citep{Guhathakurta1989}. A power-law distribution of PAH is assumed with 10 size bins \citep{Abel2008}. The dust-to-gas ratio has been linearly scaled with metallicity. 
\end{itemize}

\subsubsection{Training library with noise}\label{sec:library_noise}

The library discussed so far is made of purely theoretical models. Observed spectra contain noise which must be taken into account. Our approach is to include within the \textlcsc{ML} training phase a library containing noisy models. To this aim, we have generated a new library (100,000 models) made by two parts: the first is the original library of 50,000 photoionization models. The second is the same library to which gaussian random noise has been added to the lines with an amplitude equal to 10\% of the line intensity\footnote{Higher noise values would lead to completely noise-dominated models: given the large degeneracy in the emission lines intensities (see Sec. \ref{sec:degeneracy}), \textlcsc{GAME} would then not be able to reliably discriminate among different models.}. The sum of these two libraries represents therefore our final training dataset. We test \textlcsc{GAME} against synthetic noisy spectra in Sec. \ref{sec:noise}.

\subsubsection{Uncertainties on the inferred physical properties}\label{sec:uncertainties}

The final improvement concerns a robust estimate of the uncertainties associated with the inferred physical properties. The method works as follows. For each input spectrum, we construct $N$ modified versions of it by adding to the line intensities gaussian noise. For each input line intensity $I$ with associated error $e$, the code extracts $N$ new intensities $i$ from the following Gaussian distribution:

\begin{equation}
P(i) = \frac{1}{\sqrt{2\pi e^2}} \exp\left[-\frac{(i - I)^2}{2 e^2}\right]\,.
\end{equation}

For upper limits instead, the code generates a new line by taking a random number uniformly distributed between zero and the upper limit.

With this procedure the code generates multiple individual new observations of each spectrum. Then, for each input spectrum we obtain $N$ determinations of the quintuplet of physical properties ($n$, $N_H$, $G/G_0$, $U$, $Z$). By default the code gives as final output for each spectrum the average, the median and the standard deviation of these $N$ values. Optionally, the code can return all the $N$ determinations of the physical properties, that can be subsequently combined into a probability distribution function.

\begin{table}
	\caption{Different models used for the \textlcsc{GAME} validation test (Sec. \ref{sec:application}).}
	\centering
	\vspace{2mm}
	\begin{tabular}{c c c c c}
		\hline\hline
		name & $\log(n / {\rm cm}^{-3})$ & $\log(N_{H} / {\rm cm}^{-2})$ & $\log(U)$ & $Z / Z_{\odot}$\\
		\hline
		model (a) & 2.861 & 19.725 & -3.166 & 0.2283\\
        model (b) & 2.827 & 19.643 & -3.093 & 0.2361\\
        model (c) & 2.795 & 19.737 & -3.084 & 0.2498\\
        model (d) & 2.758 & 19.648 & -3.103 & 0.2939\\
        model (e) & 2.634 & 19.606 & -2.964 & 0.2548\\
        model (f) & 2.574 & 19.757 & -2.847 & 0.2932\\
		\hline
		\hline
	\end{tabular}
	\label{table:models}
\end{table}

\section{Validation of the code}\label{sec:application}

\subsection{Noise in observed spectra}\label{sec:noise}
We now analyse how noise on the line intensities can affect the determination of the physical properties.

We start from the emission lines of a synthetic spectrum for which the true physical properties, i.e. those used to generate it, are known. The effect of noise can be mimicked by perturbing each of the synthetic line intensities around its original value. If we then apply \textlcsc{GAME} to the perturbed intensities, it is possible to assess how the inferred physical properties vary as a function of the noise amplitude. This can be done by using in the \textlcsc{ML} training phase both the original library (50,000 models) and the one including noise (Sec. \ref{sec:library_noise}). To perform this analysis we apply the following steps:

\begin{itemize}
\item{generate two synthetic spectra: \quotes{model (a)} and \quotes{model (b)} (see Table \ref{table:models} for details);}
\item{choose a set of emission lines to use for the calculation (reported in Table \ref{table:lines_noise});}
\item{choose a set of 20 values of noise percentages ($n_i$ = 1\%, 2\%, 3\%, ..., 20\%) and compute 50 different realizations of these two spectra for each $n_i$ value: for each realization we added to the emission line intensities a gaussian random value between 0 and $n_i$ percent of the intensity of the line itself;}
\item{run \textlcsc{GAME} to infer the values of the physical properties on both the original library (50,000 photoionization models) and the noisy library (100,000 models, see Section \ref{sec:library_noise}).}
\end{itemize}

\begin{table}
	\caption{Emission lines used to compute the values of the physical properties in Section \ref{sec:noise}.}
	\centering
	\vspace{2mm}
	\begin{tabular}{c c}
		\hline\hline
		line & wavelength [\AA]\\
		\hline
		[O II] & 3726\\
        {[O II]} & 3729\\
        {[Ne III]} & 3869\\
        H$\delta$ & 4102\\
        H$\gamma$ & 4341\\
        {[O III]} & 4363\\
        H$\beta$ & 4861\\
        {[O III]} & 4959\\
        {[O III]} & 5007\\
        He I & 5876\\
        {[O I]} & 6300\\
        {[N II ]} & 6548\\
        H$\alpha$ & 6563\\
        {[N II]} & 6584\\
        {[S II]} & 6717\\
        {[S II]} & 6731\\
        {[Ar III]} & 7135\\
		\hline
		\hline
	\end{tabular}
	\label{table:lines_noise}
\end{table}

\begin{figure*}
	\centering
	\includegraphics[width=1.0\linewidth]{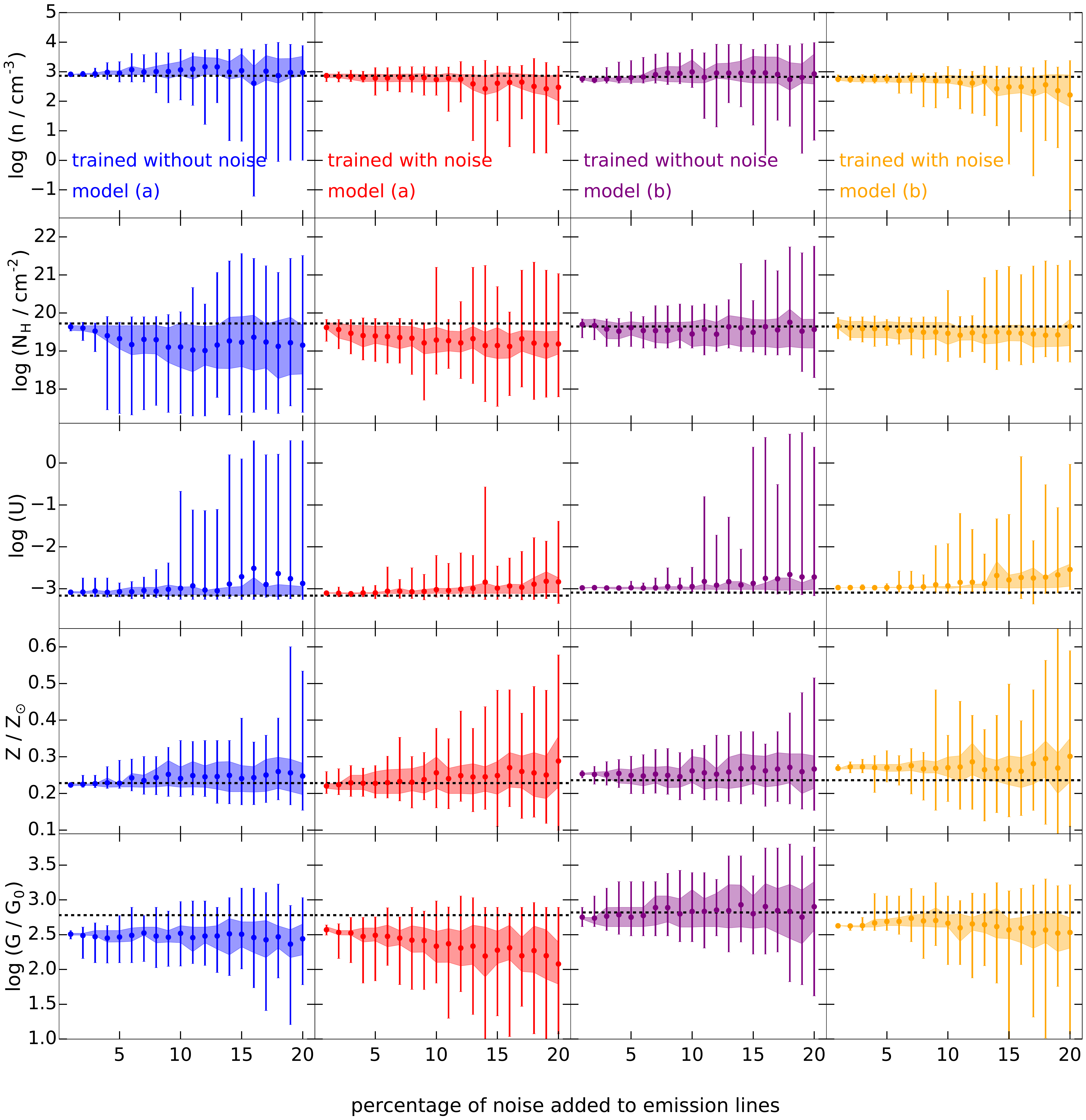}
	\caption{Inferred values of the physical properties for model (a) and (b) in Table \ref{table:models} with noise added to the emission line intensities. The circles represent the mean of the inferred values for each of the 50 realizations and the error bars denote the minimum and maximum value inferred. The shaded region represents values between the first and third quartile. The dashed horizontal lines is the \quotes{true} value used to generate the two models.}
	\label{fig:noise}
\end{figure*}

Results are shown in Figure \ref{fig:noise}, where we report the inferred physical properties for models (a) and (b). When the code is trained without noise -- first and third columns for models (a) and (b), respectively -- adding noise to the spectrum to a level as low as $n_i$ = 4\% can lead to differences between the true and inferred properties up to 2 orders of magnitude. Noisier spectra yield even larger differences.

Second and fourth columns of Figure \ref{fig:noise} show that by including the noise in the training procedure reduces these differences in the determination of the physical properties. Interestingly, we also see that metallicity determinations are quite robust. In fact, even adding noise at 20\% level the inferred metallicity is within a factor of 2 from the true value, confirming previous conclusions in \citetalias{Ucci2017}.

Although the library has been constructed with a noise percentage up to 10\%, training the \textlcsc{ML} algorithm with this library allows \textlcsc{GAME} to return outputs consistent with the \quotes{true} values even if applied to spectra with greater noise. In fact, as can be seen from Fig. \ref{fig:noise}, the mean of the inferred values agrees quite well with the \quotes{true} values up to noise level as large as 20\%.

Using noiseless libraries with low SNR spectra can lead to wrong determinations of the intrinsic physical properties (see Fig. \ref{fig:noise}). This issue is even more severe for weak lines such as [O I] $\lambda$6300 or [Ar III] $\lambda$7135 (see Table \ref{table:lines_noise}) which can be fundamental for the determination of the physical properties (i.e. they could have a very high feature importance, see Appendix \ref{sec:importance}).

\subsection{Emission lines degeneracy}\label{sec:degeneracy}

Measurements of ISM physical properties are generally based on the comparison between observations and empirically calibrated line ratio or synthetic spectra obtained from photoionization models.

\begin{figure}
	\centering
	\includegraphics[width=1.0\linewidth]{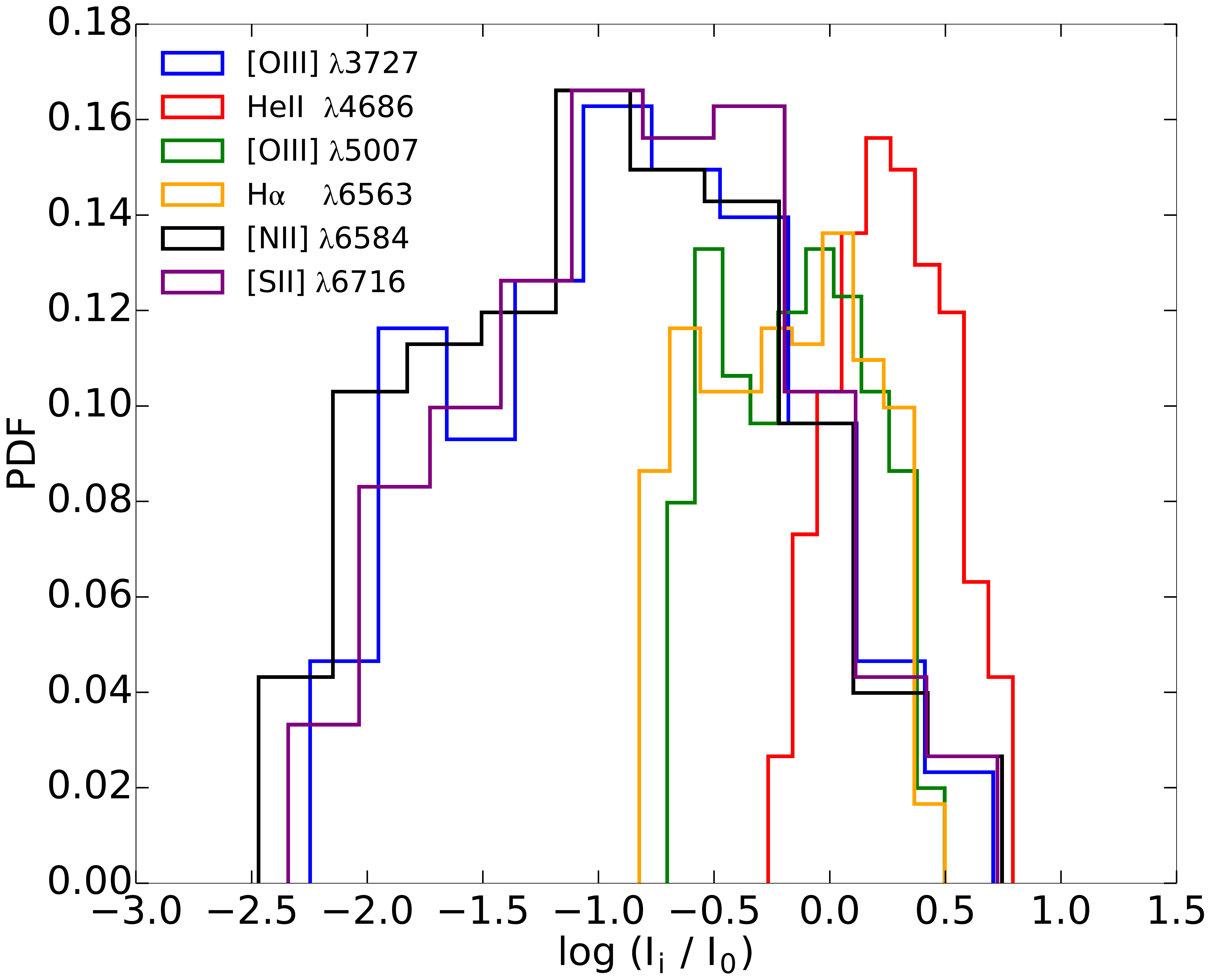}
	\caption{Ratios of the line intensities, $I_i$, for different emission lines between the i-th model and the first one (intensity $I_0$) within our set of synthetic models described in Section \ref{sec:degeneracy}.}
	\label{fig:hist_deg}
\end{figure}

\begin{figure}
	\centering
	\includegraphics[width=1.0\linewidth]{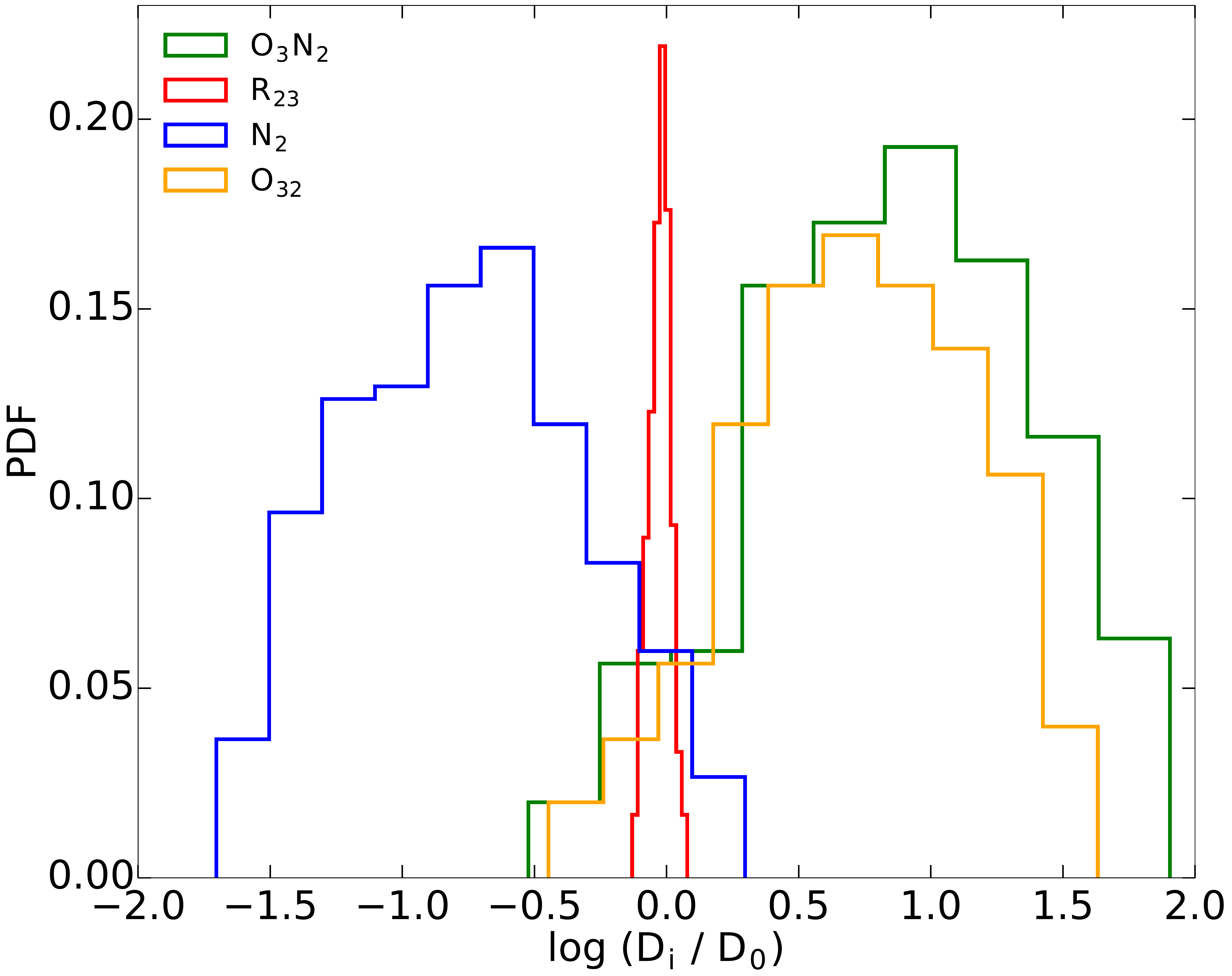}
	\caption{Ratios of commonly used line intensity diagnostics (${\rm O}_{3}{\rm N}_{2}, {\rm R}_{23}, {\rm N}_{2}, {\rm O}_{32}$) between the i-th model ($D_i$) and the first one ($D_0$) within our set of synthetic models described in Section \ref{sec:degeneracy}.}
	\label{fig:hist_deg_ratio}
\end{figure}

\begin{figure*}
	\centering
	\includegraphics[width=0.85\linewidth]{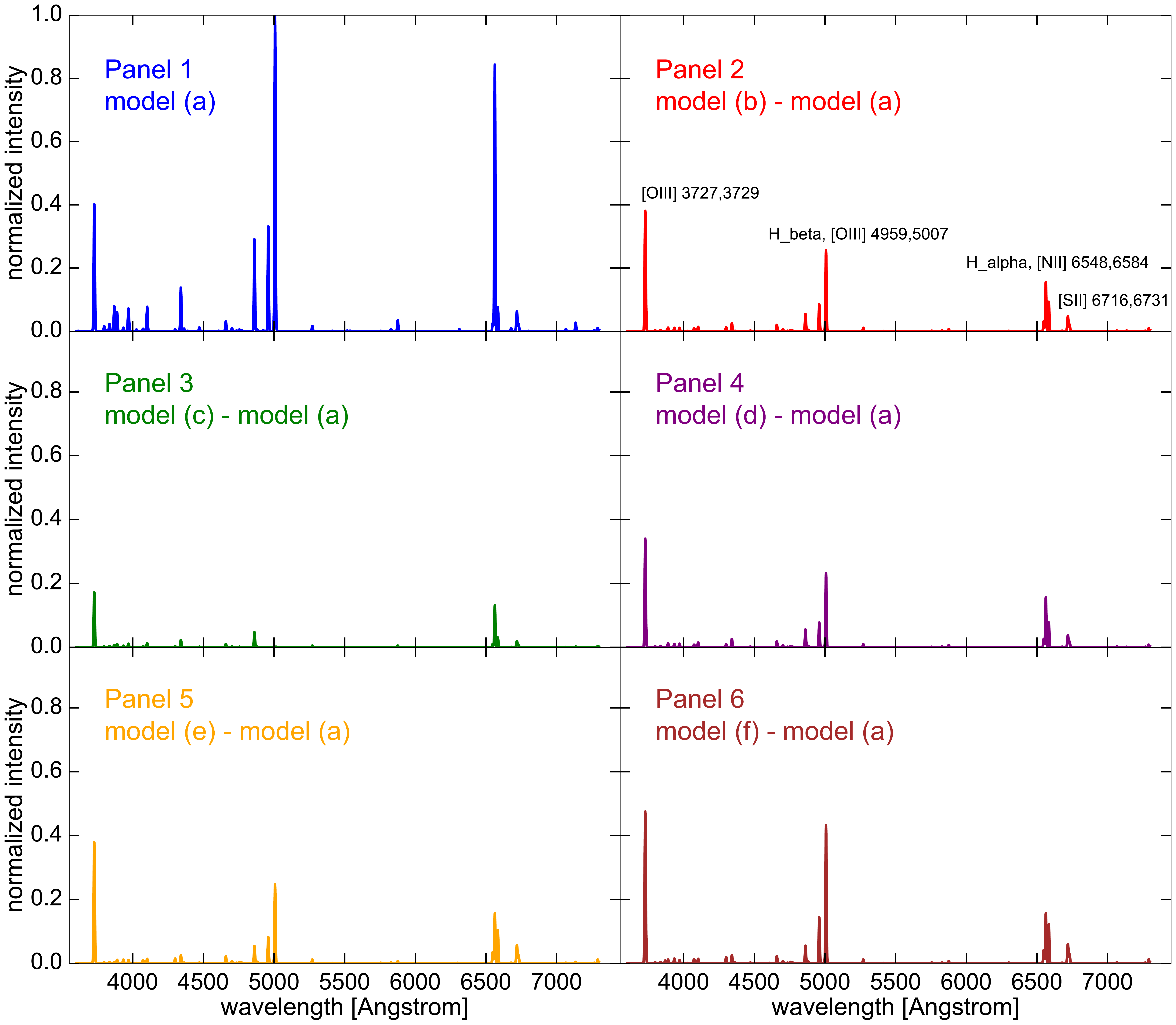}
	\caption{Optical spectra of the models reported in Table \ref{table:models}. Panel 1 shows model (a); other panels show the difference between models (b)-(f) and model (a).}
	\label{fig:spettri_vicini}
\end{figure*}

Although empirical calibrations using the electronic temperature are preferable (with respect to theoretical calibrations) because they are based on a quantity directly inferred from observables \citep{Curti2017}, they also suffer from some limitations \citep[and references therein]{Perez-Montero2017}. The major amongst these is that calibrations often use galaxy samples that do not properly cover all the physical properties space. Hence, empirical calibrations obtained from a sample of low excitation HII regions could give unreliable results when applied to global galaxy spectra \citep{Curti2017}. 

Comparisons with photoionization models are usually performed changing the metallicity and the ionization parameter but they are limited to a small range of other ISM physical properties (i.e. density) if not only to a single value \citep{Perez2014,Asari2016,Perez2017}. This is problematic, as the ISM density distribution has a dynamic range that can easily span several dex \citep{Hughes2016}.

For this reason we produced an extended library of physically motivated theoretical models (with 50,000 models described in Section \ref{sec:library}), whose purpose is to cover the large range of physical conditions found in the ISM. In this Sec. we will show that, with such a library, the emission lines arising from different combinations of the input physical properties (density, column density, ionization parameter and metallicity) are extremely degenerate even if the variation of the physical parameters is at the percent level.

We considered 300 photoionization models with the following physical properties:
\begin{align*}
-1.8 < &\log(n / {\rm cm}^{-3}) < -1.3\\
17.3 < &\log(N_H / {\rm cm}^{-2}) < 17.8\\
-1.5 < &\log(U) < -1.0\\
0.3 < &Z / Z_{\odot} < 0.6\\
\end{align*}

We then compute the following line intensities: [OIII] $\lambda$3727, HeII $\lambda$4686, [OIII] $\lambda$5007, H$\alpha$, [NII] $\lambda$6584, [S II] $\lambda$6717. In Fig. \ref{fig:hist_deg} we report the ratio of the line intensities for the i-th model and the first one within this set of 300 models. This Fig. shows that, although the physical parameters considered in the models vary within small ranges, still they result into intensity variations of several orders of magnitudes especially for [OIII] $\lambda$3727, [NII] $\lambda$6584 and [S II] $\lambda$6717. 

In Fig. \ref{fig:hist_deg_ratio}, in a similar way, we show the variation for some emission line ratios (based on common used emission line diagnostics):
\begin{align*}
{\rm O}_{3}{\rm N}_{2} &= ([{\rm OIII}]\lambda 5007/{\rm H}\beta)/([{\rm NII}]\lambda6584/{\rm H}\alpha)\,,\\
{\rm R}_{23}     &= ([{\rm OII}]\lambda 3727 + [{\rm OIII}]\lambda 4959 + [{\rm OIII}]\lambda 5007)/{\rm H}\beta\,,\\
{\rm N}_{2}      &= [{\rm NII}]\lambda 6584/{\rm H}\alpha\,,\\
{\rm O}_{32}     &= [{\rm OIII}]\lambda 5007/[{\rm OII}]\lambda 3727\,.
\end{align*}

These emission line ratios suffer also from intensity variations up to 2.5 dex. What emerges is that not only the intensity of the lines but also their ratios seem to be affected by a non negligible degeneration. Interestingly, this is not true in the case of R$_{23}$, which remains approximately constant, at least in the small range of physical properties considered in this Section. However, it must be noticed that in a larger metallicity range (-2 $\lesssim$ log(Z/Z$_{\odot}$) $\lesssim$ 0.5) R$_{23}$ is not monotonically dependent on $Z$ \citep[e.g.][first panel in Fig. 6]{Nagao2006}. Moreover, one of the fundamental advantages of using \textlcsc{GAME}, lies in the fact that instead of inferring one single physical property at one time (i.e. metallicity in the case of R$_{23}$), it can retrieve simultaneously $n$, $U$, $N_H$, $G/G_0$, $Z$ for an extended range of physical properties (see Table \ref{table:grid}).

To fully appreciate this aspect over the entire range of optical wavelengths, using the set of photoionization models listed in Tab. \ref{table:models}, we show in Fig. \ref{fig:spettri_vicini} how a small change in the physical properties is mirrored into large line intensity variations. Panel 1 of Fig. \ref{fig:spettri_vicini} shows the spectrum for model (a); the remaining panels show the spectral differences with the other models (b)-(f). Although the variation of metallicity between model (a) and (b) is $< 0.008$ dex, varying simultaneously $n$, $N_{H}$ and $U$ by 0.03, 0.08 and 0.07 dex, induces large differences in the resulting spectrum.

\textlcsc{GAME} is based on a very large library and uses all the information carried by the spectral lines on the physical properties. This approach, allowed by the \textlcsc{ML} implementation, could overcome the degeneracy better than model fitting techniques, typically based on $\chi^2$ methods, i.e. the minimization of the euclidean distance between models:

\begin{equation}
D(m,l) = \sqrt{\sum_{i=1}^{n} (m_i - l_i)^2}\,,
\label{eq:distance}
\end{equation}
where $m_i$ is the value of the emission line intensities of a given model, and $l_i$ the corresponding $i$-th value contained in the library. In the following we show that minimizing such function does not necessarily lead to a correct determination of the physical properties.

\begin{figure*}
	\centering
	\includegraphics[width=0.75\linewidth]{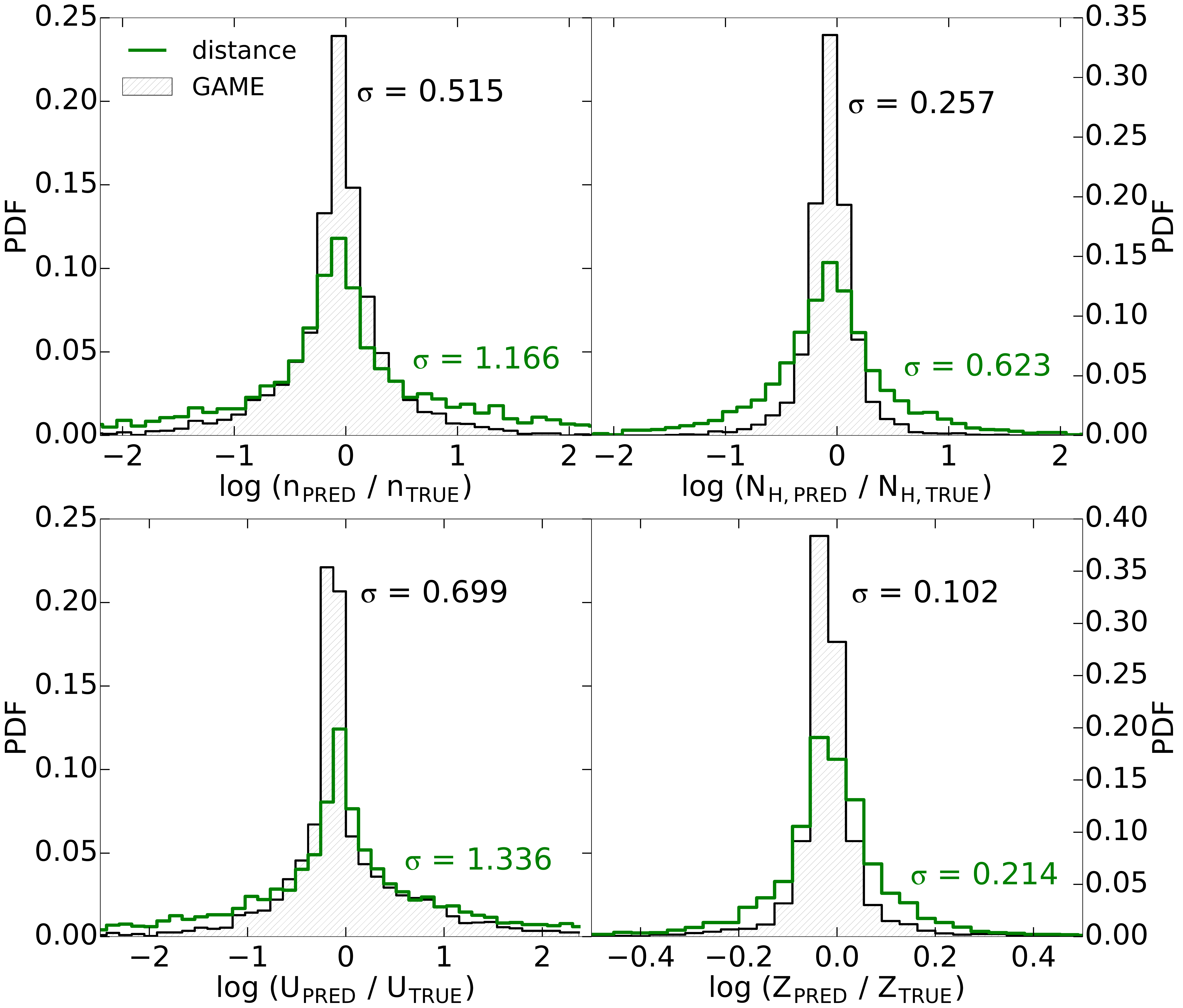}
	\caption{Ratios between the inferred and true values of the physical properties using two approaches: \textlcsc{GAME} (black shaded region) and the minimization of the euclidean distance $D(m,l)$ (green line, see equation \ref{eq:distance}). The test has been performed using 10\% of noiseless library. The standard deviations of the distributions ($\sigma$) are also shown.}
	\label{fig:nearest_ml}
\end{figure*}

\begin{table}
\caption{Set of emission lines used to asses the predictive performances of \textlcsc{GAME} (see text for details).}
	\centering
	\vspace{2mm}
	\begin{tabular}{c c}
		\hline\hline
		line & wavelength [\AA]\\
		\hline
		H$\beta$ & 4861\\
        {[O III]} & 5007\\
        He I & 5876\\
        {[O I]} & 6300\\
        H$\alpha$ & 6563\\
        {[N II]} & 6584\\
        He I & 6678\\
        {[S II]} & 6717\\
        {[S II]} & 6731\\
        \hline
		\hline
	\end{tabular}
	\label{table:lines}
\end{table}

In Fig. \ref{fig:nearest_ml} we compare the inferred values of the physical properties using the two approaches, \textlcsc{GAME} vs. $D(m,l)$ minimization (eq. \ref{eq:distance}), using the set of emission lines in Table \ref{table:lines}. We perform the test on 10\% of the noiseless library (\textlcsc{GAME} uses the remaining 90\% as the training dataset), including 5,000 models. On this reduced set we both infer the physical values with \textlcsc{GAME} and by distance minimization.

Trying to recover a spectrum that is as similar as possible to the input spectrum (minimizing the distance) leads to worse results with respect to \textlcsc{GAME}. In Fig. \ref{fig:nearest_ml} it is evident that the standard deviation of the logarithm of the ratio between the \quotes{predicted} and \quotes{true} values in the case of \textlcsc{GAME} is a factor of two smaller than the $D(m,l)$ minimization approach. The similarity between two spectra, given the extreme degeneration (see Sec. \ref{sec:degeneracy}), does not mean necessarily a good correlation with their physical properties. This was expected since other \textlcsc{ML} techniques as k-Nearest Neighbour (with k=1 in the case of equation \ref{eq:distance}) work well when the input space has a low dimensionality, which does not apply to our problem.

\section{Comparison with other methods}\label{sec:comparison}

\subsection{Overview}

The most prominent feature of \textlcsc{GAME} is that it exploits the full information encoded in a spectrum. Instead of using small, pre-selected subsets of emission line ratios, \textlcsc{GAME} can use an arbitrary number of lines to infer the ISM physical properties. Additionally, its usage is not limited to a specific ISM phase (i.e. HII regions).

Another key advantage of the \textlcsc{ML} implementation is that, once trained, it requires a very short computational time (see also Appendix \ref{sec:time}). This is a crucial feature in view of applications to modern IFU observations with $\sim$ 100,000 spaxels, each one with a substantial number (> 10) of observed lines per spectrum.

\subsection{Quantitative comparison: galaxies}

Here we quantitatively compare \textlcsc{GAME} with other empirical calibrations and results from different codes. To perform this comparison we use a sample of 62 stacked galaxy spectra. These spectra are obtained from SDSS star forming galaxies in $0.027 < z < 0.25$, and are fully described in \citet{Curti2017}. The spectra are stacked in bins of 0.1 dex according to the log values of their [OII]$\lambda$3727/H$\beta$ and [OIII]$\lambda$5007/H$\beta$ ratios. For each of the stacked spectra \citet{Curti2017} derived the corresponding metallicity, using a set of new and self-consistent empirical calibrations. We compare such determination with the results of \textlcsc{GAME} and the results from two widely used emission line codes, \textlcsc{pyqz} and \textlcsc{HII-CHI-mistry}.

\textlcsc{pyqz}\footnote{\url{http://fpavogt.github.io/pyqz/index.html}} \citep{Dopita2013} is a public Python code which uses abundance- and excitation-sensitive line ratios to define a plane in which the oxygen abundance and ionization parameter can be determined by interpolating a grid of photo-ionization models to match the observed line ratios.

\textlcsc{HII-CHI-mistry}\footnote{\url{http://www.iaa.es/~epm/HII-CHI-mistry.html}} \citep{Perez2014} is publicly-available Python code. This code takes the extinction-corrected emission line fluxes and, based on a $\chi^2$ minimization on a photoionization models grid, determines chemical-abundances (O/H, N/O) and ionization parameter. In this work we use the version 3.0 of the code, dealing with input line uncertainties via a Monte Carlo approach.

\begin{figure}
	\centering
	\includegraphics[width=0.95\linewidth]{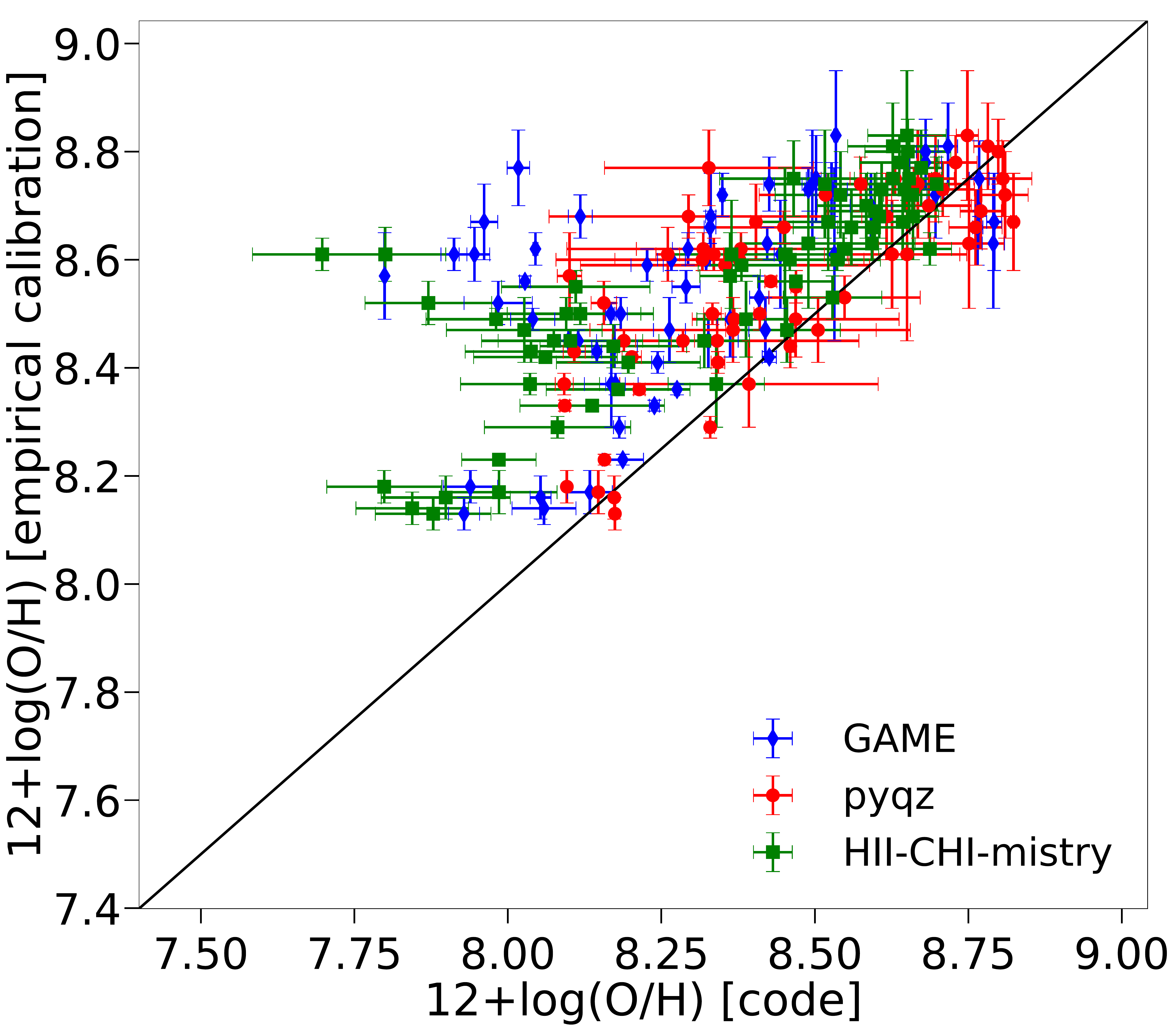}
	\caption{Comparison between the metallicity inferred with the empirical calibrators of \citet{Curti2017} and by \textlcsc{GAME} (blue diamonds), \textlcsc{pyqz} (red circles), and \textlcsc{HII-CHI-mistry} (green squares) on 62 stacked galaxy spectra (see text for details).}
	\label{fig:game_vs_method}
\end{figure}

As evident from the comparison shown in Fig. \ref{fig:game_vs_method}, the empirical method tends to slightly overestimate metallicity with respect to all codes. Besides, the mean offset and the standard deviation between \textlcsc{GAME} predictions and those from other methods are small (less than 0.3 dex). Overall, the agreement on metallicity is good.

\begin{figure}
	\centering
	\includegraphics[width=0.95\linewidth]{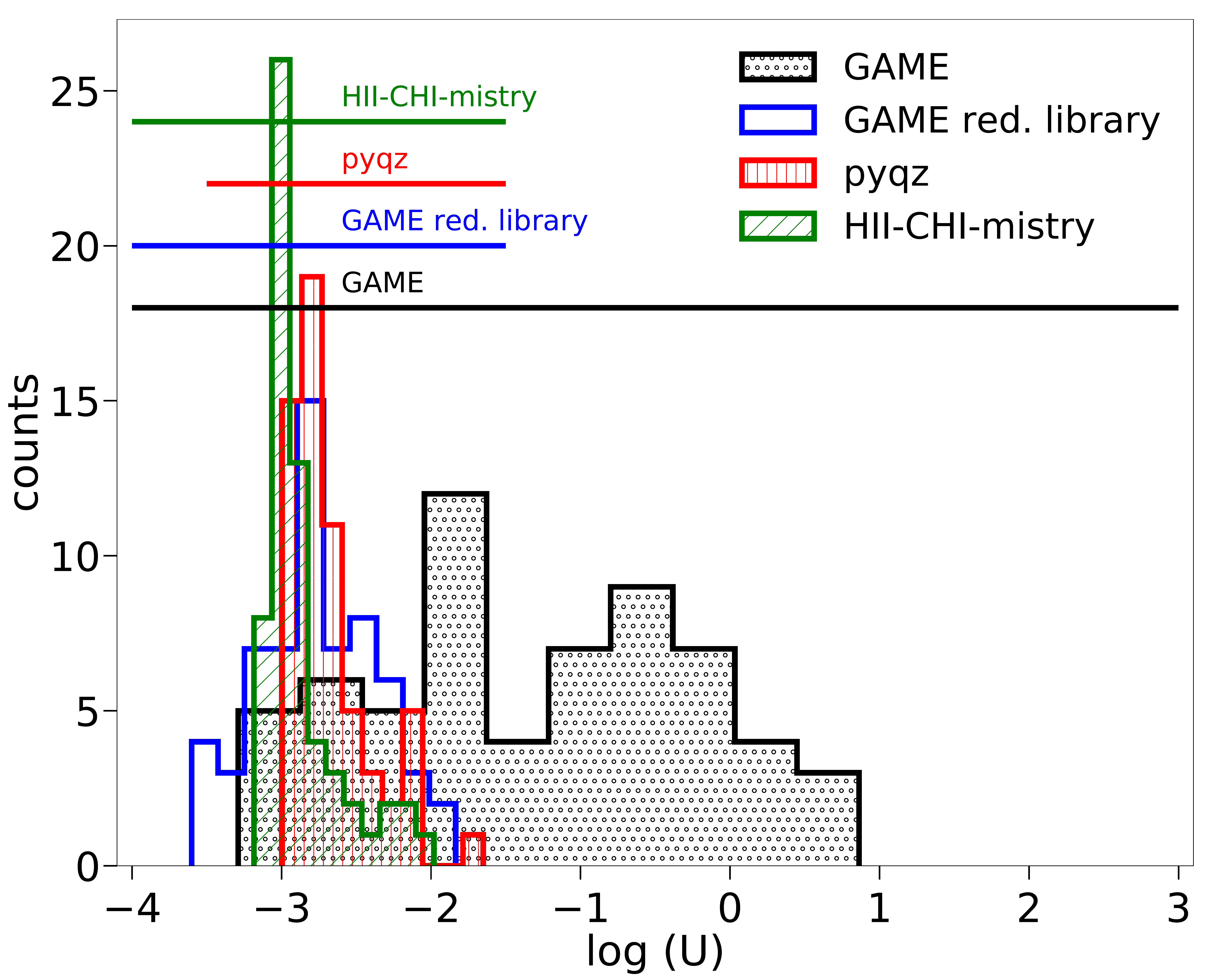}
	\caption{Ionization parameter, $U$, distribution for 62 stacked galaxy spectra inferred by \textlcsc{GAME} (grey dotted histogram), \textlcsc{GAME} with a reduced library (empty), \textlcsc{pyqz} (red), and \textlcsc{HII-CHI-mistry} (green). The horizontal lines denote the $U$ range considered in the libraries of the different codes.}
	\label{fig:histogram_U}
\end{figure}

We extend the comparison among the three codes to the ionization parameter (Fig. \ref{fig:histogram_U}). Interestingly, the \textlcsc{GAME} results markedly differ from those obtained with the other two codes. This can be due to the fact that \textlcsc{pyqz} and \textlcsc{HII-CHI-mistry} only contain models with a narrower $U$ range (i.e. -3.5 < log $U_{\textlcsc{pyqz}}$ < -1.5, and -4.0 < log $U_{\textlcsc{HII-CHI-mistry}}$ < -1.5). To verify this point, we ran \textlcsc{GAME} with a reduced library containing ionization parameter values in the range -4.0 < log($U$) < -1.5. Fig. \ref{fig:histogram_U} clearly shows that with the reduced library \textlcsc{GAME} infers $U$ values that are in agreement with those from the other two codes, e.g. the mean of the inferred values are respectively $\mu$(\textlcsc{pyqz}) = -2.67, $\mu$(\textlcsc{GAME}) = -2.76, $\mu$(\textlcsc{HII-CHI-mistry}) = -2.88. Hence it is important to consider $U$ values that are larger than usually assumed.

Values of $U$ up to 10 are expected in line-emitting galaxies. Let us consider the relation between the ionizing photon flux and the star formation rate \citep[e.g.][]{Murray2010}:

\begin{equation}
\text{Q(H)} = 2.46 \times 10^{53}\left(\frac{\text{SFR}}{\text{M}_{\odot}\text{yr}^{-1}}\right)  \text{s}^{-1}
\end{equation}

For a compact galaxy with a SFR = 1 M$_{\odot}$ yr$^{-1}$, $n$ = 1 cm$^{-3}$, and radius 1 kpc, we obtain log($U$) $\sim$ 0.6, consistently with the upper limit found with \textlcsc{GAME} (see Fig. \ref{fig:histogram_U}). As discussed in \citetalias{Ucci2017}, the spectrum emerging from a galaxy is weighted by the column densities along the line of sight. If diffuse phases are dominant in the build-up of the final spectrum, their low density pushes $U \propto n^{-1}$ towards large values. 

Note that although the full library contains spectra with -4.0 < log($U$) < 3 (see Table \ref{table:grid}), \textlcsc{GAME} does not infer values log($U$) > 1. This implies that this bound is set by the physical conditions, and it is independent of the library extension.

We finally note that $U$ is the physical parameter most affected by uncertainties. The bootstrap routine now included in the code should however significantly mitigate the problem: although the inferred values for U cover a large range, their PDF obtained with the bootstrap is highly peaked (see Appendix \ref{sec:boot} and Fig. \ref{fig:game_pdf}).

\subsection{Quantitative comparison: HII regions}\label{sec:hii}

As an additional comparison, we apply \textlcsc{GAME} to a sample of 75 observed HII regions with available chemical abundance determinations. We choose HII regions for which a large number ($N>13$) of high quality (typical errors <10\%) emission lines is available. Our final sample is composed by:

\begin{itemize}
\item 66 HII regions located in 21 dwarf irregular galaxies observed by \citet{vanZee2006}. For 25 HII regions, oxygen abundances are obtained via direct detection of emission lines tracing the electron temperature; for the remaining 41 HII regions, abundances are inferred through SEL methods. The emission lines (where available) used as input for \textlcsc{GAME} are: [OII] $\lambda\lambda$3727+3729, [NeIII] $\lambda$3869, [OIII] $\lambda\lambda$4959+5007, [OI] $\lambda$6300, [SIII] $\lambda$6312, H$\alpha$, [NII] $\lambda\lambda$6548+6584, HeI $\lambda$6678, [SII] $\lambda\lambda$6717+6731, [ArIII] $\lambda$7135 (Table 3 of that paper);

\item 9 HII regions located towards the Galactic anti-centre observed by \citet{Fernandez2017} for which more than 60 emission lines measurements are available (Table A.1 of that paper).
\end{itemize}

\begin{figure}
	\centering
	\includegraphics[width=0.90\linewidth]{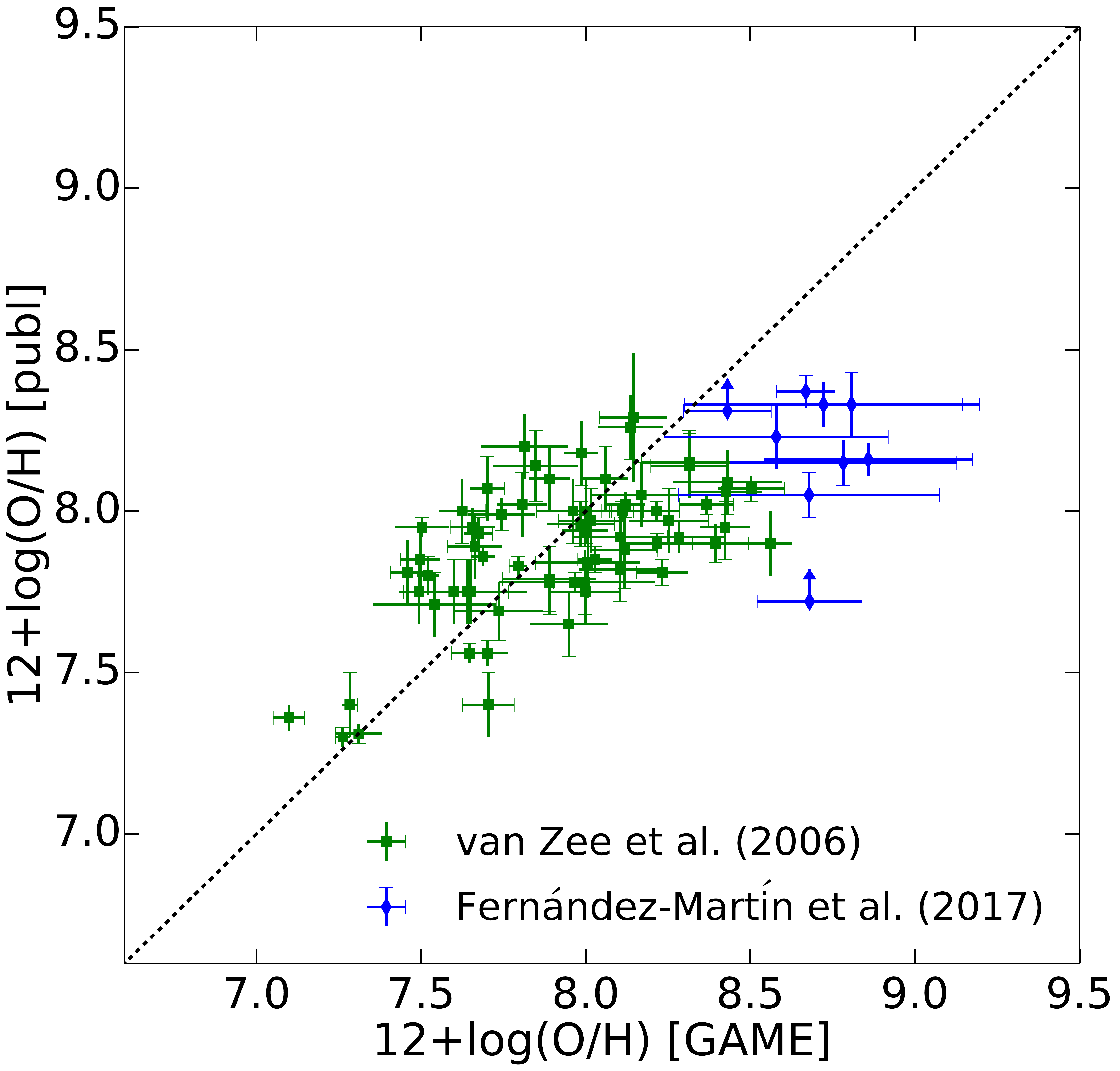}
	\caption{Comparison of the oxygen abundance determinations for HII regions reported in \citet{vanZee2006} and \citet{Fernandez2017} with those inferred by \textlcsc{GAME}}.
	\label{fig:hii}
\end{figure}

Results are shown in Fig. \ref{fig:hii}, where we compare the abundance determinations reported in the cited works with the \textlcsc{GAME} results. The metallicity values inferred by \textlcsc{GAME} ($Z/Z_{\odot}$) are converted into oxygen using the relation 12+log(O/H) = 8.69. The overall agreement between the two methods is good. The scatter does not seem to correlate with metallicity, and the dispersion is of $\simeq 0.2 - 0.3$ dex. The mean offset (between the results inferred with \textlcsc{GAME} and the published results reported in Fig. \ref{fig:hii}) and its standard deviation are respectively 0.08 and 0.29. Differences among the results of different methods can arise because of several reasons. First of all, the oxygen abundance we infer from the metallicity assumes solar abundances. Moreover, while \textlcsc{GAME} adopts input emission lines that are not de-reddened, the reddening corrections adopted by \citet{vanZee2006} and \citet{Fernandez2017} (i.e. $n_e$ = 100 cm$^{-3}$, Case B recombination, different extinction curves) may have a significant impact on the physical properties determination.

The main advantage of \textlcsc{GAME} is that it does not use any assumption on the gas physical properties (e.g. density, temperature) defining the emission line spectrum we are looking at. The code is able to recover the physical properties (in this case metallicity) by extracting them from a library containing a vast collection of physical conditions of the ISM. We also note that \textlcsc{GAME} does not even use the information that the sample refers to HII regions: it simply searches the library, trains itself, and gives the best predicted physical conditions.

\section{Summary and discussion}\label{sec:discussion}

We presented an updated and optimized implementation of \textlcsc{GAME} \citepalias{Ucci2017}, a code designed to infer ISM physical properties from emission line spectra. The code is based on a \textlcsc{ML} algorithm (AdaBoost with Decision Trees as base learner) to calculate density ($n$), column density ($N_H$), ionization parameter ($U$), metallicity ($Z$) and FUV flux in the Habing band ($G/G_0$) given an arbitrary set of emission line flux measurements (or upper limits) with their uncertainties.

\textlcsc{GAME} is extremely reliable particularly for the metallicity determination: the 5-fold cross-validation score with the set of emission lines reported in Table \ref{table:lines} is higher than 0.95. Although some properties, as metallicity, are easily and robustly recovered also from noisy spectra with few emission lines, other physical properties require higher quality data with more lines measurements. We have shown that for a given set of emission lines if the cross-validation score for the metallicity is 0.95, it might happen that lower scores are obtained for $n$ and $N_H$ (both $\sim$ 0.8) and $U$ ($\sim$ 0.7).

Another key point in our analysis is that the emission lines intensities are highly degenerate. A small variation of the physical properties leads to large changes in the emission line intensities ratios. The \textlcsc{ML} approach used here can overcome this issue much better than classical fitting methods based on $\chi^2$ minimization. Noticeably, such important result can be achieved also when the spectra include noise as in real observations.

We have compared \textlcsc{GAME} with methods based on empirical calibrations \citep{Curti2017} and other codes (\textlcsc{pyqz} and \textlcsc{HII-CHI-mistry}) by considering a sample of 62 stacked spectra from SDSS galaxies \citep{Curti2017}. While a very good agreement has been obtained in terms of the metallicity determination for the considered sample, we find discrepancies in the derived values of the ionization parameter. We discuss possible reasons for such disagreement in Sec. \ref{sec:comparison}.

Finally, we have tested our code on a sample of 75 HII regions with direct method and Strong Emission Lines (SEL) abundances determinations \citep{vanZee2006,Fernandez2017} to study how \textlcsc{GAME} can recover these values. We found that the oxygen abundances determinations are in good agreement with those inferred with \textlcsc{GAME} with a typical scatter around 0.2 - 0.3 dex. The applications of \textlcsc{GAME} are not only limited to HII regions but the code can deal equally well with different phases of the ISM, including the molecular one. This is because the underlying library covers the largest possible range of physical conditions characterizing the ISM. Furthermore, \textlcsc{GAME} offers the possibility to use an arbitrary set of emission lines which span a wavelength range from the Ly$\alpha$ one (1216 \AA) to 1 mm. These features allow the user to infer the physical properties of phases ranging from the Hot Ionized Medium (HIM) to dense molecular cores.

One of the main limitations of \textlcsc{GAME} (and other methods/codes) relies on the possible presence of different ISM phases/gradients along the line of sight contributing to the same spectrum. This introduces a complexity that cannot be managed at the present time. We preliminarily discussed this issue in \citetalias{Ucci2017}. The main result there was that in such conditions the returned physical parameters are biased towards the phase with largest gas column density. To make progress, we plan to investigate this issue in more details using emission line spectra generated from high-resolution galaxy simulations. Simulated galaxies and their spectra offer the advantage that the physical conditions of the gas shaping the observed spectra is precisely known. This will allow us to (a) devise more stringent reliability tests for \textlcsc{GAME}, and (b) understand how to maximize the information retrieval from spectra arising from multiphase lines of sight (for this issue we also refer the reader to Section 5.3 of \citetalias{Ucci2017}).

Is is nowadays possible to obtain spatially resolved spectra of galaxies. The information coming from different regions of a galaxy requires the development of new methods in order to obtain the physical conditions of the different phases of the ISM. Large libraries and robust algorithms will be crucial in the analysis of galaxy spectra which include faint lines arising from extended wavelength ranges. Combining the UV/optical/IR/FIR information from the same object will be the next step towards a better understanding of the internal structure of distant galaxies.

\section*{Acknowledgements}
We thank M. Curti for providing the galaxy spectra used in our analysis and the anonymous referee for constructive insights. We also thank B. Greig, N. Gillet, C. Behrens, and L. Vallini for useful discussions and comments. AF acknowledges support from the ERC Advanced Grant INTERSTELLAR H2020/740120

\bibliographystyle{mnras}
\bibliography{bibliography}

\appendix

\section{Workflow of the code}\label{sec:work}

\begin{figure*}
	\centering
	\includegraphics[width=1.0\linewidth]{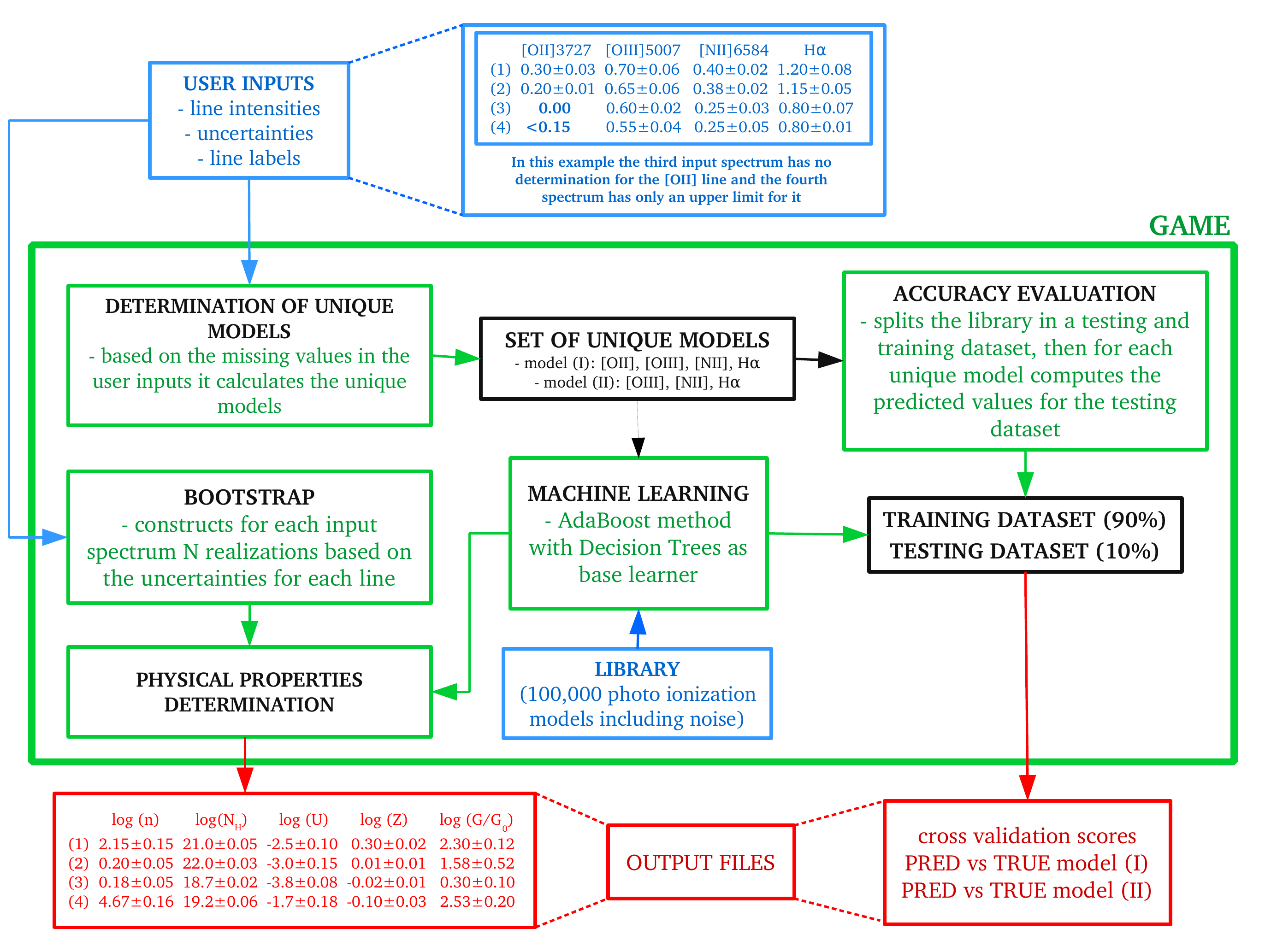}
	\caption{\textlcsc{GAME} workflow diagram showing the main routines (green boxes). Execution requires: (a) user inputs, and (b) a library of photoionization models (blue boxes). The main output files (red boxes) are (a) the physical properties with associated uncertainties, and (b) the cross-validation scores. The input spectra and the physical property values are shown purely for illustration.}
	\label{fig:work_flow}
\end{figure*}

Fig. \ref{fig:work_flow} shows \textlcsc{GAME} workflow diagram. The diagram presents also an example where 4 input spectra with a maximum of 4 emission lines are used to infer physical properties \footnote{We recall that the inferred physical properties are 5, but only 4 are used to generate the \textlcsc{Cloudy} models: $n$, $N_H$, $U$, $Z$.}. We underline that \textlcsc{GAME} can deal with missing lines (third spectrum) or upper limits (fourth). We will refer to this example throughout the Section. The input spectra and the physical property values shown are purely for illustration. 

\subsection{Determination of unique models routine}\label{sec:unique}
As the AdaBoost \textlcsc{ML} algorithm does not deal with missing values, the approach we followed is to simply get rid of emission lines for which a measurement (or at least an upper limit) is not available. The first routine therefore computes the number of unique models (i.e. unique combinations of input emission lines) from the input file. In the example in Fig. \ref{fig:work_flow} there are two different unique models: (I) one including all the 4 input lines, and (II) another including all but the [OII] line. Each unique model requires a separate  \textlcsc{ML} training phase. 

\subsection{Accuracy evaluation routine}
The goal of any \textlcsc{ML} model, is to learn from the training set and generalize for unseen data. The accuracy is a measure of the performance on unseen data. To quantify the accuracy, we split the spectral library into a training (90\%) and testing (10\%) dataset. The \quotes{accuracy evaluation routine} trains the \textlcsc{ML} algorithm on the training dataset, and predicts values for the testing one. The predicted values are then compared to the actual values used to generate the testing spectra.

In addition, we define a variable $k$ that splits the whole library into $k$ equal parts called \quotes{folds}. The routine then performs the so-called $k$-fold cross-validation in which the \textlcsc{ML} is trained on $k$-1 folds and tested on the remaining fold. 

\begin{figure}
	\centering
	\includegraphics[width=0.95\linewidth]{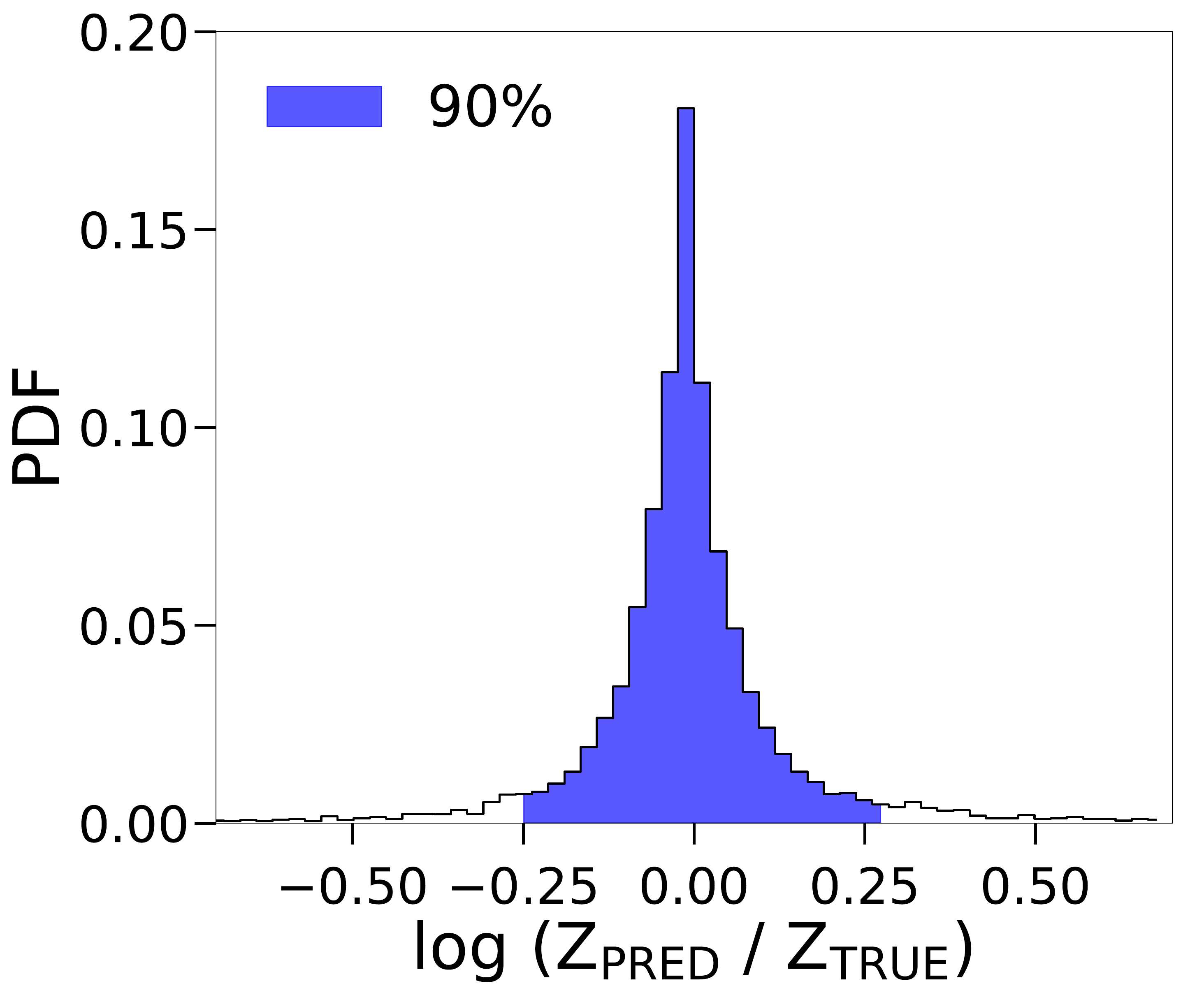}
	\caption{Predicted/true ratio distribution for metallicity values using the set of emission lines in Table \ref{table:lines_appendix}. 90\% of the models reside within the shaded blue region.}
	\label{fig:pred_vs_true}
\end{figure}

For each $k$ fold, let $\bar{p}_k$ be the mean of the inferred physical properties:
\begin{equation}
\bar{p}_k = \frac{1}{n} \sum_{i=1}^{n} p_{i,k}
\end{equation}
where $n$ = (length of library) / $k$. We then define the two following quantities:

\begin{equation}
T_k = \sum_{i=1}^{n} (p_{i,k} - \bar{p}_k)^2 ,
\end{equation}

\begin{equation}
R_k = \sum_{i=1}^{n} (p_{i,k} - t_{i,k})^2 ,
\end{equation}
where $t_{i,k}$ are the true values. For each $k$ and physical property the routine computes the coefficient of determination,

\begin{equation}
R^2_{k} = 1 - \frac{R_k}{T_k}.
\end{equation}
The final output is the mean score $R^2 =\langle R^2(k)\rangle$ and its standard deviation. If the predicted values perfectly fit the true data, $R^2=1$. Thus $R^2$ represents a measure of the accuracy of the method.   

Figure \ref{fig:pred_vs_true} shows the predicted/true ratio distribution for metallicity values using the set of emission lines in Table \ref{table:lines_appendix}. About 90\% of the models are within a factor of 2 from the true values, yielding $R^2>0.95$ (5-fold cross-validation). This is a remarkable performance as it has been obtained with a relatively small number of emission lines. 
\begin{table}
\caption{Emission lines used to quantify \textlcsc{GAME} accuracy.}
	\centering
	\vspace{2mm}
	\begin{tabular}{c c}
		\hline\hline
		line & wavelength [\AA]\\
		\hline
		H$\beta$ & 4861\\
        {[O III]} & 5007\\
        He I & 5876\\
        {[O I]} & 6300\\
        H$\alpha$ & 6563\\
        {[N II]} & 6584\\
        He I & 6678\\
        {[S II]} & 6717\\
        {[S II]} & 6731\\
        \hline
		\hline
	\end{tabular}
	\label{table:lines_appendix}
\end{table}
\subsection{Machine Learning Routine}
This routine trains the AdaBoost \textlcsc{ML} on the whole library (the one including noise: 100,000 models) and predicts the values for the physical properties. For the example in Fig. \ref{fig:work_flow} the \textlcsc{ML} routine is called only twice, one for each of the two unique models (I) and (II). Since there are 5 physical properties to be determined ($n$, $N_{H}$, $G/G_0$, $U$, $Z$), the training procedure is composed by 5 different sub-trainings. We therefore use for the AdaBoost algorithm each label separately and independently.

\subsection{Bootstrap routine}\label{sec:boot}
To understand the effects of noise in observed spectra we have devised the following procedure. For each input library spectrum this routine builds $N$ modified versions of it by adding gaussian noise (see Sec. \ref{sec:uncertainties}) to the line intensities. 
\begin{figure*}
	\centering
	\includegraphics[width=0.8\linewidth]{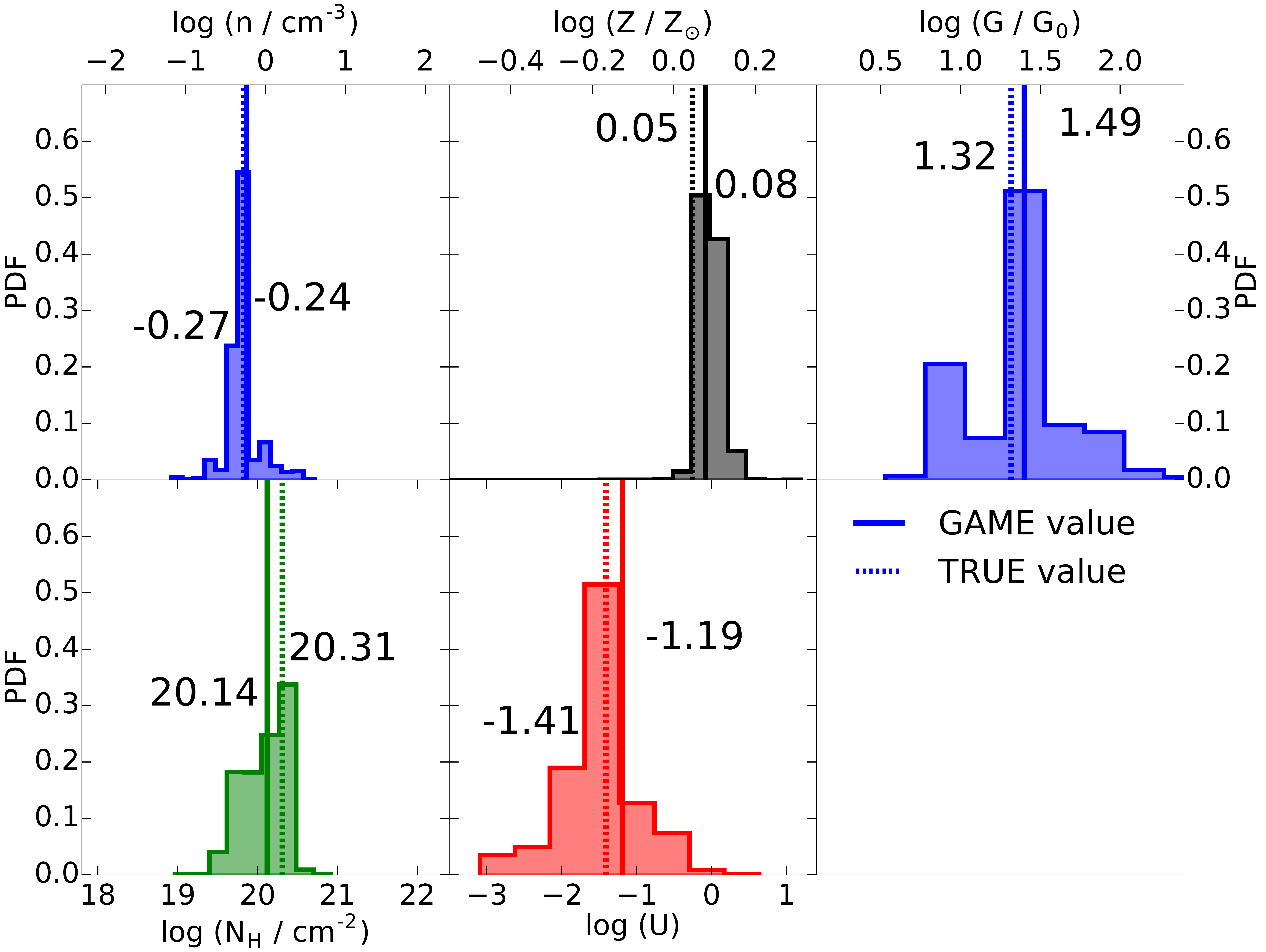}
	\caption{Distribution of the inferred physical properties using $N = 10,000$ realizations of a single synthetic input spectrum. Vertical lines denote the true (dashed) and GAME (solid) values whose values are also shown.}
	\label{fig:game_pdf}
\end{figure*}
We started from a synthetic spectrum featuring the emission lines in Table \ref{table:intensities} (\quotes{intrinsic} values). Then we add gaussian noise with r.m.s. amplitude equal to 15\% of the line intensity (\quotes{error} values in Table \ref{table:intensities}).
Fig. \ref{fig:game_pdf} shows the distributions for the inferred physical properties of $N$ = 10,000 realizations\footnote{We have verified that the distributions do not change appreciably for larger $N$ values.} for this synthetic input spectrum.

\begin{table*}
	\caption{Physical properties for the model used in the analysis of Section \ref{sec:boot}. We report both the true and predicted values using the mean of $N$=10,000 realizations. Also reported are the intrinsic values of the emission line intensities relative to H$\alpha$ and their associated errors.}
	\centering
	\vspace{2mm}
	\begin{tabular}{c c c c c c c c c c}
		\hline
        \hline
		 & $\log(n / {\rm cm}^{-3})$ & $\log(N_{H} / {\rm cm}^{-2})$ & $\log(U)$ & $Z / Z_{\odot}$ & $\log(G / {\rm G}_{0})$ \\
		\hline
		true & -0.271 & 20.309 & -1.411 & 1.1106 & 1.318\\
        predicted & -0.239 & 20.142 & -1.190 & 1.2023 & 1.489\\
        \hline
        \hline
         & H$\beta$ & {[O III]} 5007 & He I 5876 & {[O I]} 6300 & H$\alpha$ & {[N II]} 6584 & He I 6678 & {[S II]} 6717 & {[S II]} 6731\\
        \hline
        intrinsic & 0.2931 & 0.0286 & 0.0402 & 0.0045 & 1.0000 & 0.0706 & 0.0119 & 0.0550 & 0.0374\\
        error & 0.0156 & 0.0040 & 0.0026 & 0.0003 & 0.0215 & 0.0020 & 0.0001 & 0.0007 & 0.0020\\
        \hline
		\hline
	\end{tabular}
	\label{table:intensities}
\end{table*}

Metallicity is the most accurately-inferred property showing the narrowest distribution (maximum width $\sim$ 0.2 dex). The inferred and true $Z$ values differ only by 0.02 dex. Although the distributions of the other physical properties are broader, the inferred mean value of, e.g. $U$, differs only by 0.22 dex from the true one.

\section{Feature importance}\label{sec:importance}
In most cases input features do not contribute equally to predict the outputs. In many situations a certain fraction of the features could be also effectively irrelevant.

The notion of \quotes{feature importance} measures the frequency\footnote{Here for frequency we mean the ratio between the number of occurrences of a feature as split point and the total number of split points in the Decision Tree.} of the usage of each feature as split point in a tree, and the resulting \quotes{purity}\footnote{In order to decide which feature to split on at each step in building the tree, the algorithm chooses the split that results in the purest daughter nodes. The purity is usually quantified by the reduction of the mean squared error or the entropy.} of the branches determined by a feature separation. Feature importances are defined for the specific \textlcsc{ML} algorithm used, and the specific set of input lines.

For individual Decision Trees the feature importance interpretation is relatively easy because it is possible to simply look at the tree structure. In fact, a tree intrinsically performs a feature selection by selecting appropriate split points. For the AdaBoost method, which could include several trees, the interpretation becomes more difficult as feature importance is defined as the mean over all trees.

\textlcsc{GAME} determines the feature importances as real numbers in [0,1] (the feature importances are normalized so that they sum up to 1). The higher the value, the more important is the contribution of the feature to the accuracy. The implementation in the code uses the Gini importance \citep{Breiman2001,Breiman2002}. \footnote{ More details on the features importance applied to the AdaBoost technique can be found in \citet{Hoyle2015}.} Gini importance, sometimes called also \quotes{mean decrease impurity}, is defined as the total decrease in node impurity (weighted by the probability of reaching that node) averaged over all trees of the ensemble \citep[see also][]{Louppe2013}. 

\begin{figure}
	\centering
	\includegraphics[width=1.0\linewidth]{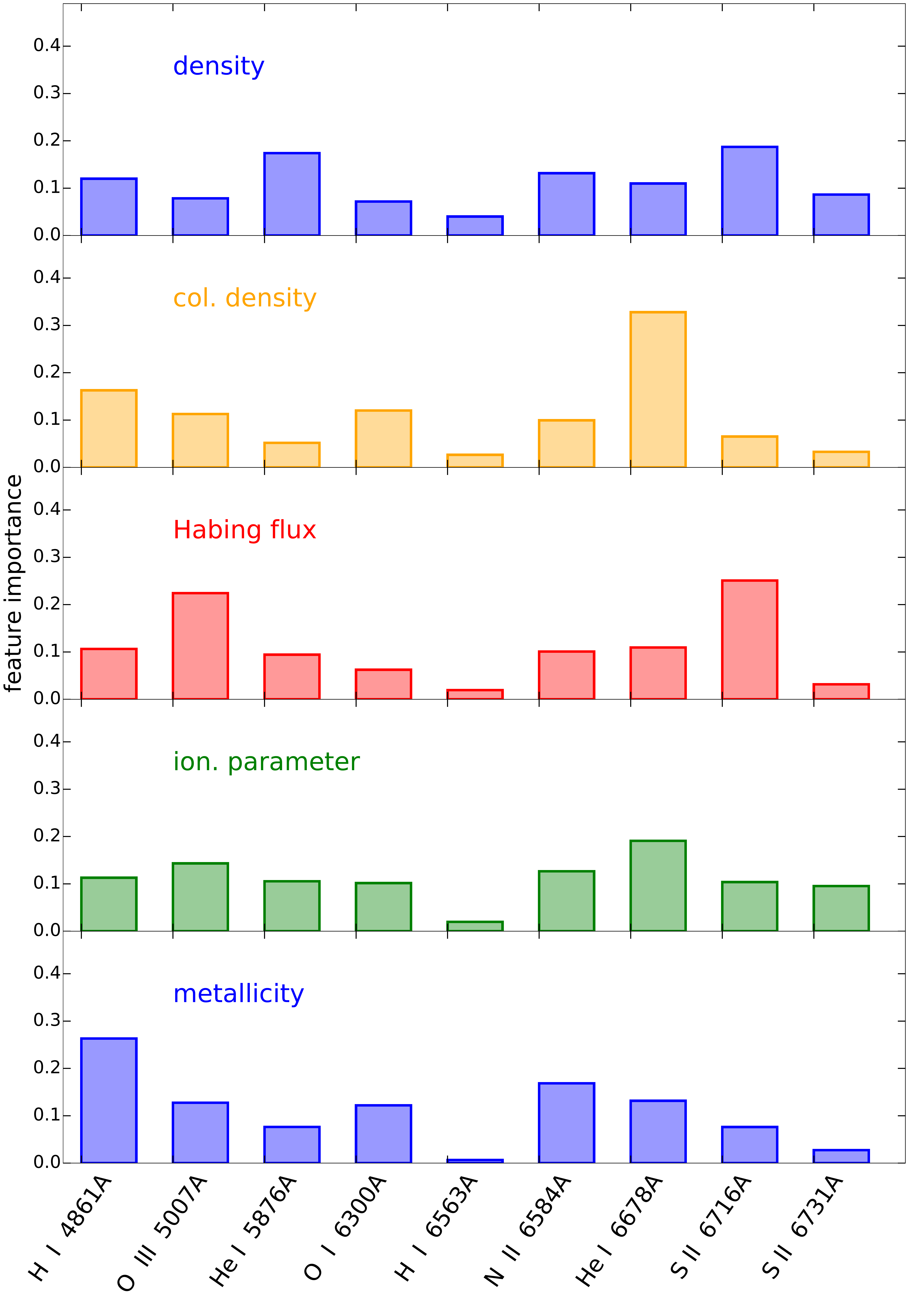}
	\caption{Feature importance for different physical properties for the AdaBoost method applied to the emission lines in Tab. \ref{table:lines_appendix}.}
	\label{fig:importances}
\end{figure}

As an example, we show in Fig. \ref{fig:importances} the feature importance for a subset of emission lines. For all the physical properties considered, the H$\alpha$ line turns out to be the least important feature. Perhaps surprisingly, we find that the metallicity determination is most affected by the H$\beta$ line. Its importance is $\sim 2$ times larger than that of the other emission lines.

From the above, it should be clear that feature importance notion is algorithm-dependent, and does not represent a direct one-to-one connection with a specific physical property. This is because the feature importance of a given line results from the combination of photoionization physics and the architecture of the Decision Trees. For this reason, a purely physical interpretation of such indicator might be misleading.

Nevertheless, the feature importance provides a first indication of the optimal set of lines to be considered as an input for GAME to recover a physical property of interest. For example, lines with high feature importance, if present in the observed spectrum, should be included in the ML analysis as they improve the accuracy of the results. This criterion applies only for lines with a high signal-to-noise ratio, as otherwise the associated noise would limit the benefit of their inclusion. 

\section{Execution time}\label{sec:time}
The total wall-clock time (on a CPU Intel Core 2 QUAD CPU Q9550/2.83GHz) for running \textlcsc{GAME} on the synthetic spectrum of Section \ref{sec:boot} with the full library of 100,000 models that includes noise, is around 14 minutes. The sole execution time of the routines that involve the \textlcsc{ML} algorithm is instead about 13 minutes. These performances have been obtained setting the number of $k$-folds for the cross-validation score equal to 5, and the total number of trees to 50.

We stress that almost all the running time is used up by the training procedure. If the observed spectra all contain the same lines (i.e. only one unique model) the most time-consuming phase needs to be performed only once. Thus, the total running time will be essentially independent of the number of spectra. 

\section{Accuracy }\label{sec:appendix_b}
We finally discuss how the accuracy of the results depends on the number of lines used and their feature importance. 

\begin{figure}
	\centering
	\includegraphics[width=0.95\linewidth]{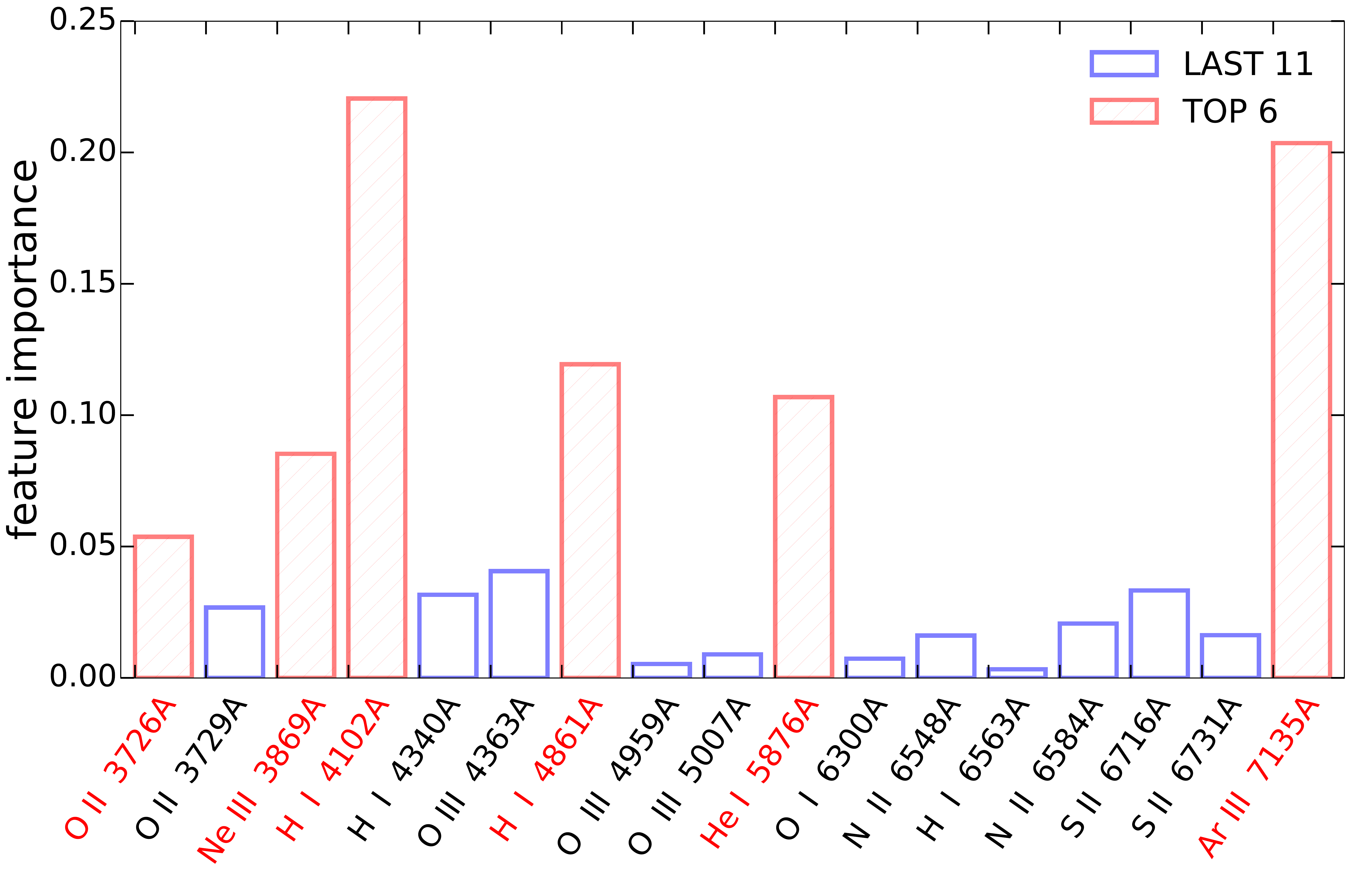}
	\caption{Feature importance for the metallicity determination and  the set of emission lines reported in Table \ref{table:lines_noise}. The shaded (empty) boxes report the 6 most (least) informative features.}
	\label{fig:importances_appendix}
\end{figure}

We start by showing (Fig. \ref{fig:importances_appendix}) the feature importance of the 17 lines listed in Table \ref{table:lines_noise} for the metallicity label. As in Appendix \ref{sec:importance}, the H$\alpha$ line turns out to be the least important feature. The 6 most informative features are lines connected to H and He transitions (i.e. H$\beta$, H$\delta$, HeI) and 3 metals lines (i.e. [OII], [NeIII], and [ArIII]). Note also that the two most informative lines, H$\delta$ and [ArIII], sum up to $>40\%$ of the total feature importance.

Next, we have performed two \textlcsc{GAME} runs. In run A we analysed 16 different models, with an increasing number of input line intensities (from $l=2$ to $l=17$) ordered by their wavelength. Run B is identical, but the lines are added sequentially according to their feature importance (Fig. \ref{fig:importances_appendix}).

\begin{figure*}
	\centering
	\includegraphics[width=0.76\linewidth]{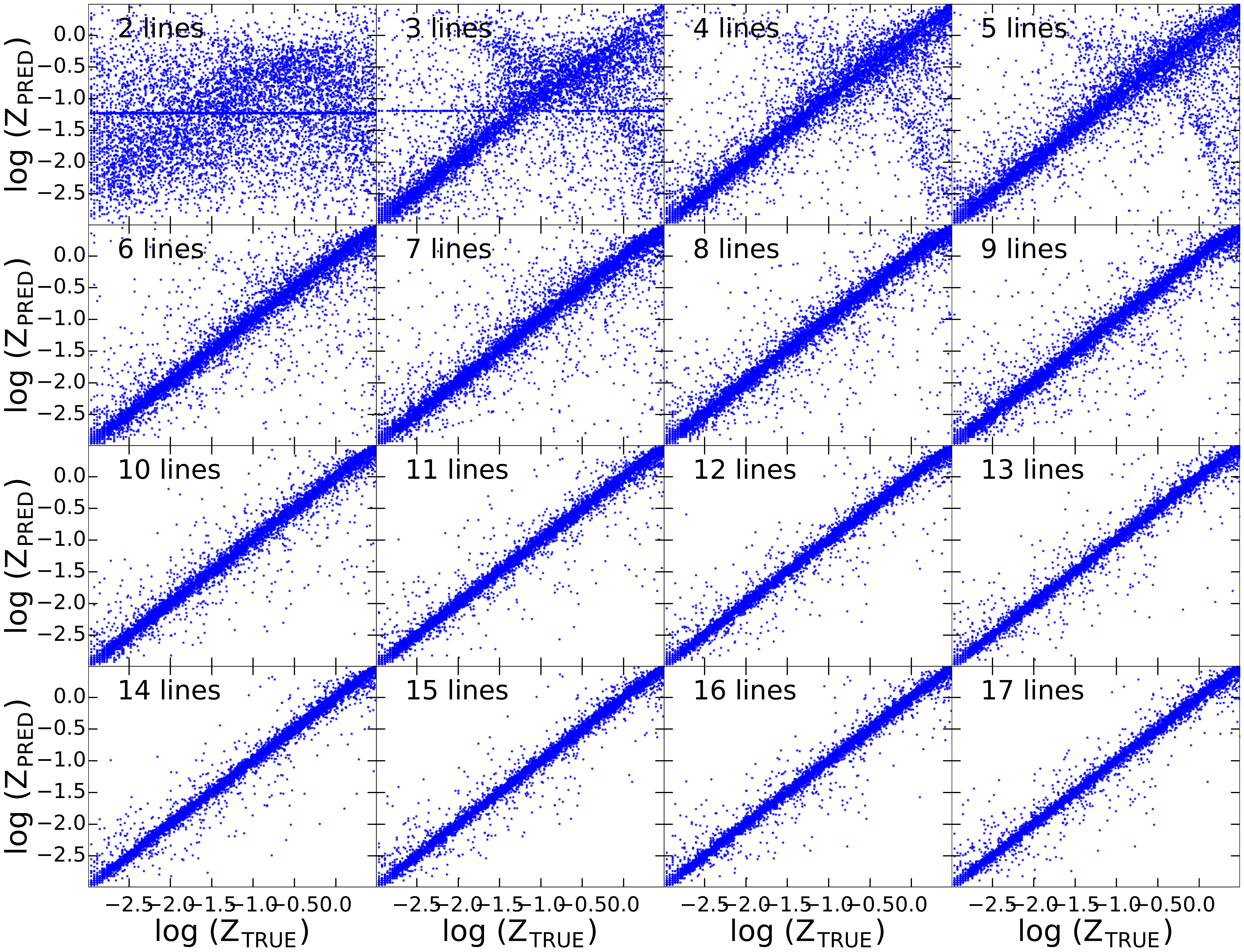}
	\caption{Predicted vs. true metallicities in the testing dataset. Panels refer to runs with a different number of input lines ordered by their wavelength (run A).}
	\label{fig:true_vs_true_lambda}
\end{figure*}

\begin{figure*}
	\centering
	\includegraphics[width=0.74\linewidth]{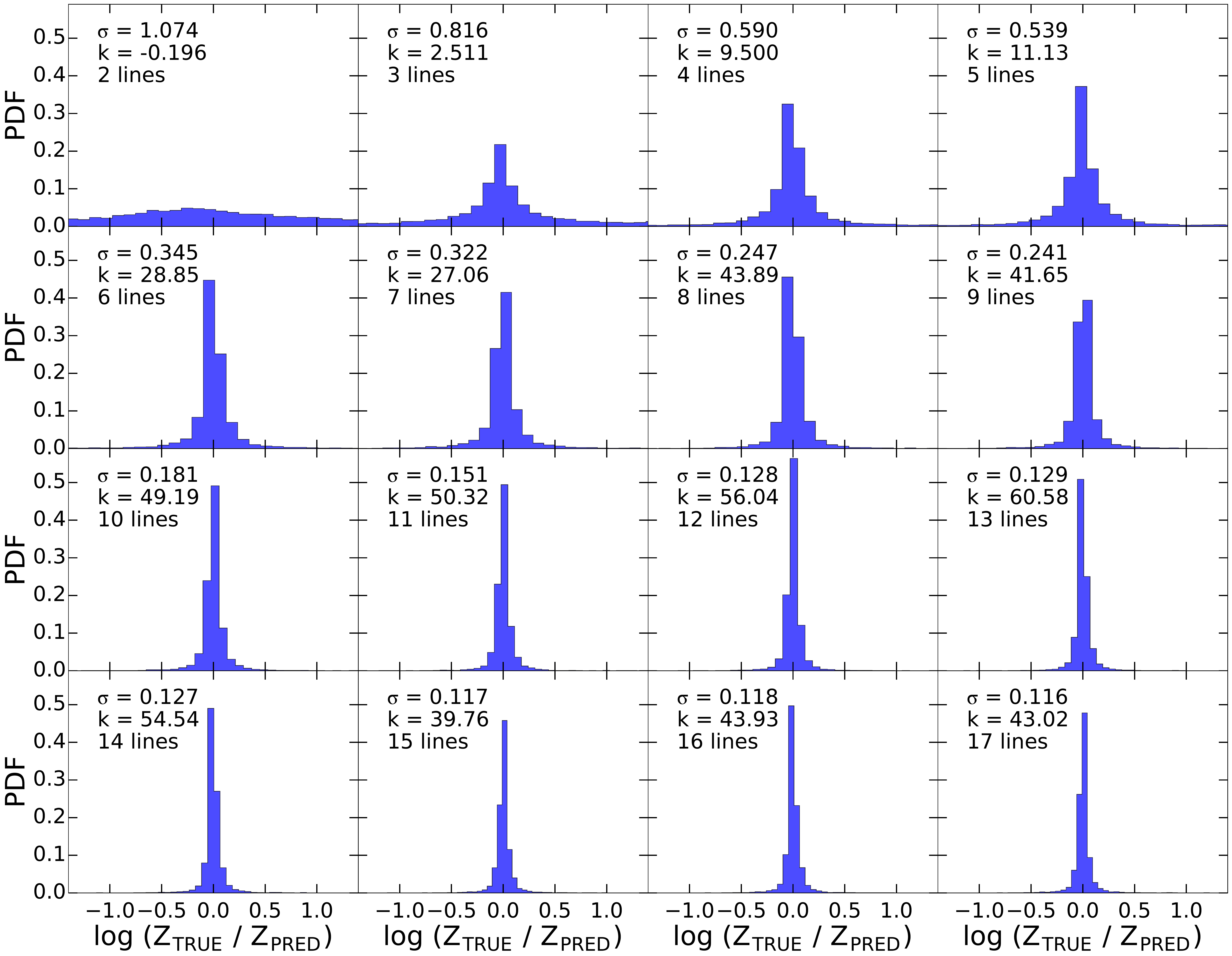}
	\caption{Probability distribution functions (PDF) of the predicted/true metallicity ratio in the testing dataset. 
Panels refer to runs with a different number of input lines ordered by their wavelength (run A). $\sigma$ and $k$ denote the standard deviation and kurtosis, respectively.}
	\label{fig:hist_lambda}
\end{figure*}

\begin{figure*}
	\centering
	\includegraphics[width=0.76\linewidth]{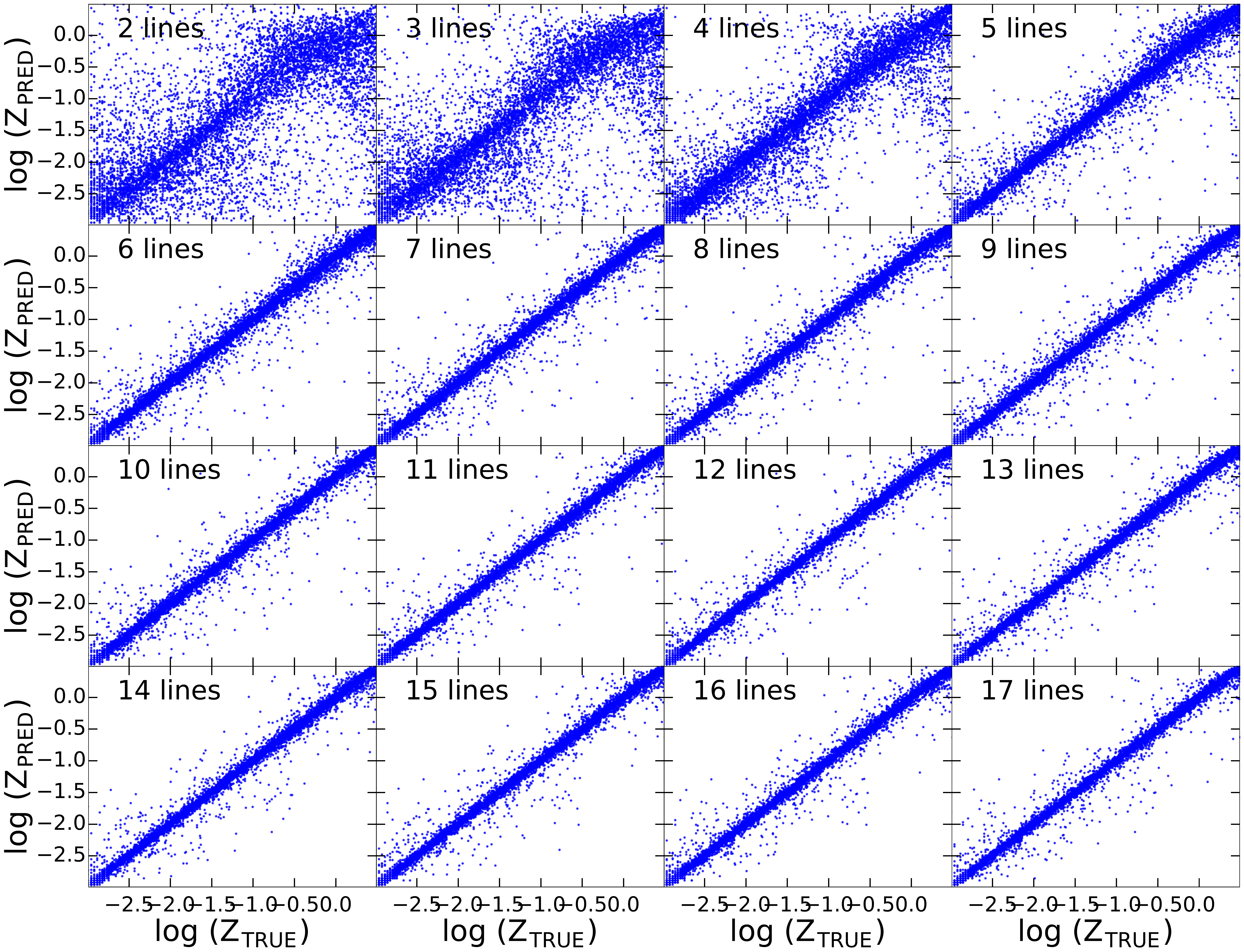}
	\caption{Predicted vs. true metallicities in the testing dataset. Panels refer to runs with a different number of input lines ordered by their feature importance (run B).}
	\label{fig:true_vs_true_feat}
\end{figure*}

\begin{figure*}
	\centering
	\includegraphics[width=0.74\linewidth]{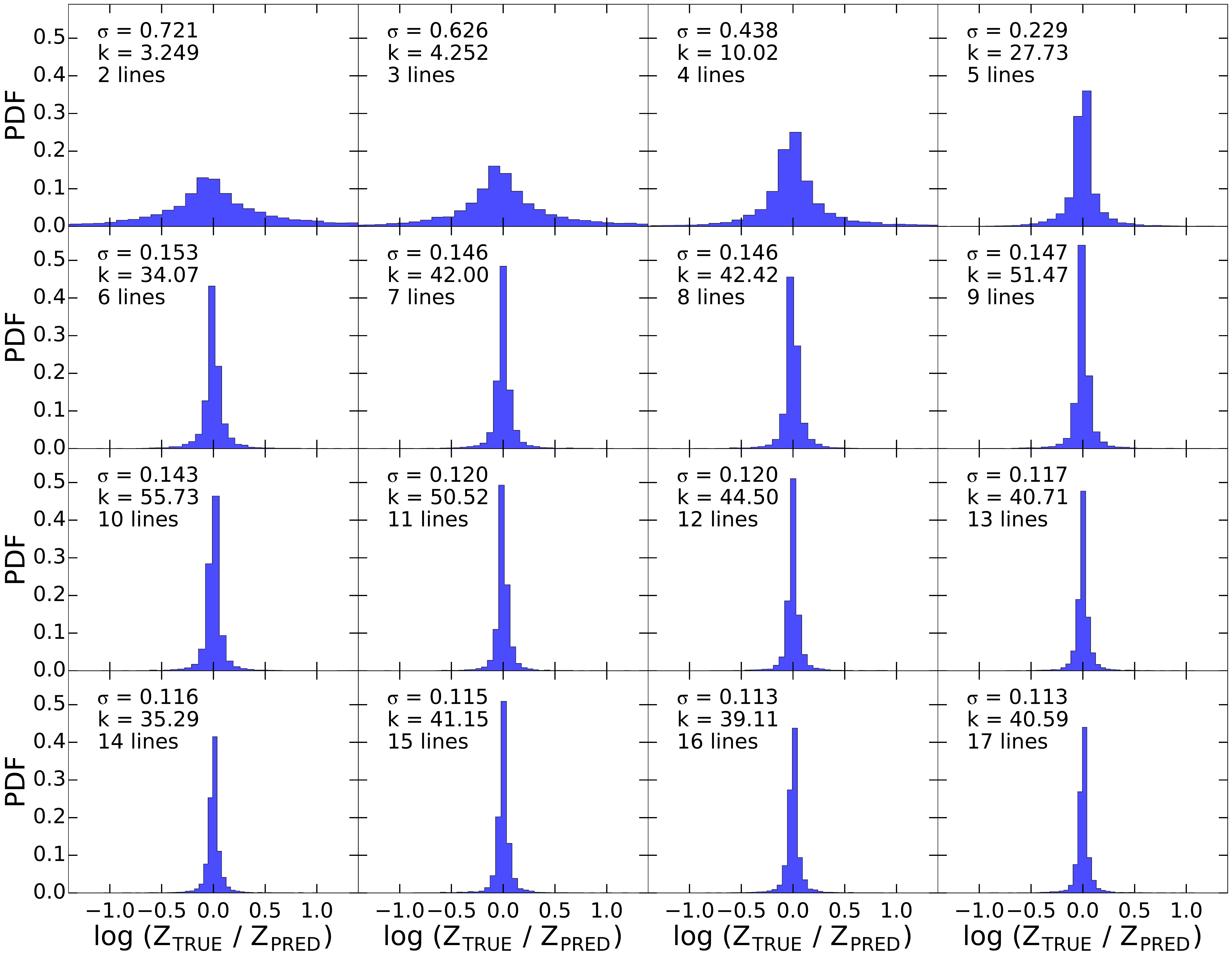}
	\caption{Probability distribution functions (PDF) of the predicted/true metallicity ratio in the testing dataset. 
Panels refer to runs with a different number of input lines ordered by their feature importance (run B). $\sigma$ and $k$ denote the standard deviation and kurtosis, respectively.}
	\label{fig:hist_feat} 
\end{figure*}

\begin{figure}
	\centering
	\includegraphics[width=0.85\linewidth]{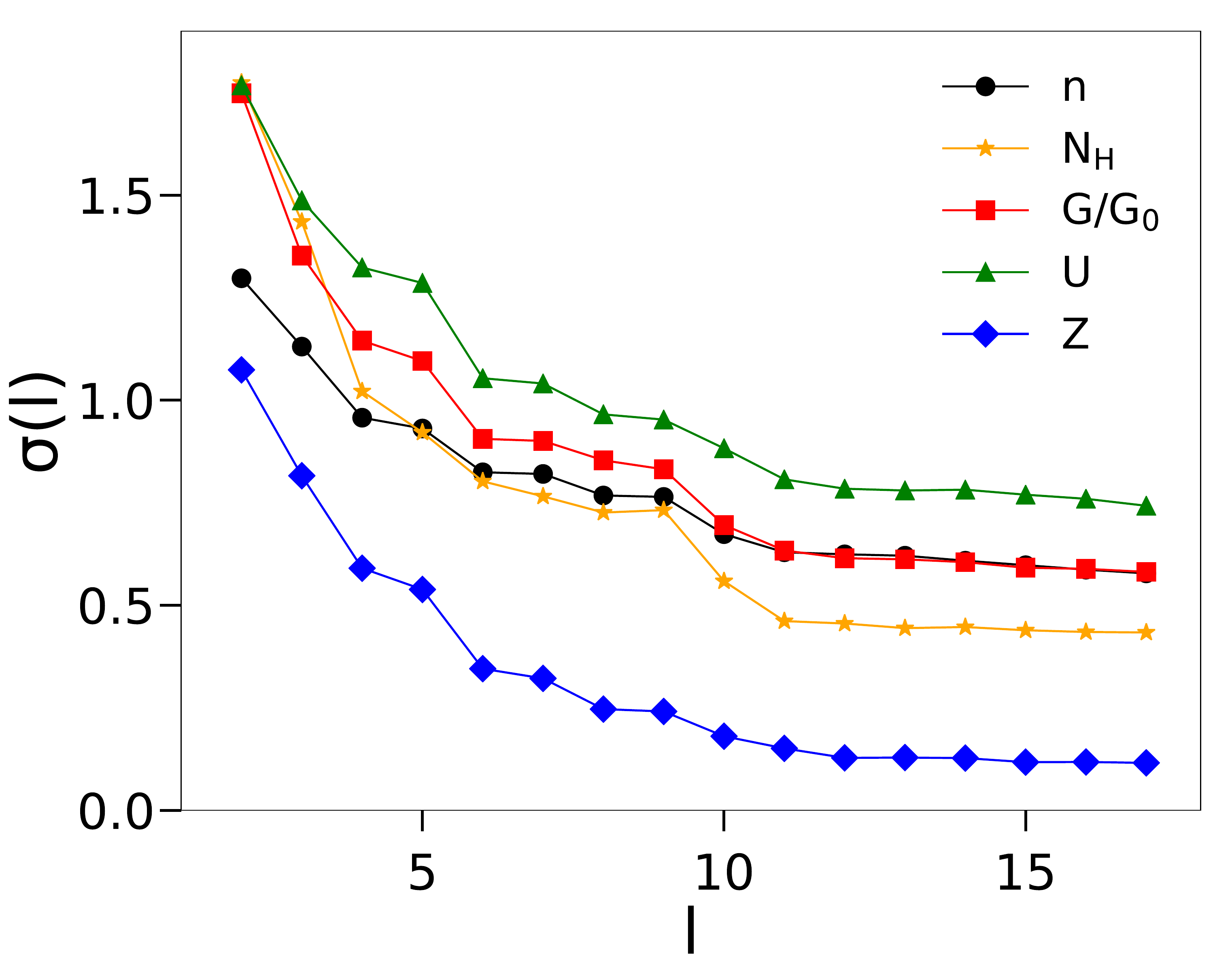}
	\caption{Standard deviation of models with $l$ input lines. }
	\label{fig:sigmas}
\end{figure}

The results of the metallicity in run A are reported in Figs. \ref{fig:true_vs_true_lambda} and \ref{fig:hist_lambda}. Although it is technically possible to construct Decision Trees even with only two lines, a huge scatter is present in the first two panels of Fig. \ref{fig:true_vs_true_lambda}. Also noticeable is a thick horizontal line in the distribution, i.e. in many cases \textlcsc{GAME} predicts the same value. This implies that the code is not effectively learning from data. As a consequence the PDF of the inferred metallicity values for a small number of lines (Fig. \ref{fig:hist_lambda}) is essentially flat.

Figs. \ref{fig:true_vs_true_feat} and \ref{fig:hist_feat} illustrate\footnote{The careful reader might have noticed that the standard deviation, $\sigma$, and kurtosis, $k$, of the PDF with 17 input lines is slightly different from those given in Fig. \ref{fig:hist_lambda}. This is because changing the order of the input features in the boosted Decision Trees may result in slightly different Trees.}  the results of run B. Contrary to the previous case, \textlcsc{GAME} starts to learn already with just two lines, and the scattering remains always smaller than in run A case for any number of input lines. Selecting the input lines according to their feature importance produces clear advantages: using the most important 6 lines produces an accuracy improvement of a factor 2.25 over the wavelength-ordered addition. 

In both runs, the fraction of data outliers drastically drops as $l$ is increased from $l=2$ to $l=10$. Beyond that value the standard deviation flattens out to a small value, $\sigma \approx 0.1$. To further check this conclusion, We have performed the same study also for all the labels considered here ($n$, $N_{H}$, $U$, $Z$, $G/G_0$). The trend described above is indeed common to all these labels, as it can be appreciated from Fig. \ref{fig:sigmas}, where we plot the standard deviation as a function of the number of lines used.

In the example of Fig. \ref{fig:work_flow}, we use 4 input emission lines to infer 5 physical properties, out of which 4 are independent. In the context of traditional statistics, the number of degrees of freedom of the problem is dof = \# of input emission lines - \# of physical independent properties $= 0$.

In this Sec. however, we have inferred physical properties using less then 4 emission lines (Figs. \ref{fig:true_vs_true_lambda} and \ref{fig:true_vs_true_feat}). \textlcsc{ML} and traditional statistics are based on different approaches. \textlcsc{ML} does not assume any parametric function for describing the data. The method is instead empirical: the outputs or even the applicability of the model are checked a posteriori. Thus, it is not straightforward to extend the dof definition to a \textlcsc{ML} algorithm. \textlcsc{GAME} can build trees even with a negative number of dof; it just requires $n \geq 2$ lines. A posteriori we can see that only when the code uses more than $n \gtrsim 6$ lines the results are statistically reliable (Figs. \ref{fig:hist_lambda} and \ref{fig:hist_feat}).

\bsp	
\label{lastpage}
\end{document}